\documentclass[fontsize=6pt,journal=nalefd,manuscript=letter]{achemso}
\usepackage{graphicx} 
\usepackage[utf8]{inputenc}
\usepackage{amsmath}
\usepackage[dvipsnames]{xcolor}
\usepackage[numbers,sort&compress,super]{natbib}
\usepackage{caption}
\usepackage{hyperref}
\usepackage{siunitx}
\usepackage{soul}
\usepackage{gensymb}
\usepackage{caption}
\usepackage{subcaption}
\usepackage{newtxtext,newtxmath}
\usepackage{mleftright}
\usepackage{mathtools}
\setkeys{acs}{articletitle = true}

\DeclarePairedDelimiter\bra{\langle}{\rvert}
\DeclarePairedDelimiter\ket{\lvert}{\rangle}
\DeclarePairedDelimiterX\braket[2]{\langle}{\rangle}{#1 \delimsize\vert #2}

\usepackage{bm}

\newcommand{\expect}[3]{\ensuremath{\left<{#1}\left|{#2}\right|{#3}\right>}} 
\newcommand{\eqn}[1]{\begin{equation}{#1}\end{equation}}

\newcommand\blfootnote[1]{%
  \begingroup
  \renewcommand\thefootnote{}\footnote{#1}%
  \addtocounter{footnote}{-1}%
  \endgroup
}

\title{Strong coupling between localized surface plasmons and molecules by coupled cluster theory} 
\author{$^\Delta$Jacopo Fregoni}
\affiliation{Dipartimento di Scienze Chimiche, University of Padova, I-35131 Padova, Italy}
\alsoaffiliation{Istitute of Nanosciences, Consiglio Nazionale delle Ricerche CNR-Nano, I-41125 Modena, Italy}
\author{$^\Delta$Tor S. Haugland}
\affiliation{Department of Chemistry, Norwegian University of Science and Technology, 7491 Trondheim, Norway}
\author{Silvio Pipolo}
\affiliation{UCCS Unit\'e de Catalyse et Chimie du Solide, Universit\'e de Lille, Universit\'e d’Artois UMR 8181, F-59000, Lille, France}
\author{Tommaso Giovannini}
\affiliation{Scuola Normale Superiore, I-56126, Pisa, Italy}
\author{Henrik Koch}
\affiliation{Scuola Normale Superiore, I-56126, Pisa, Italy}
\alsoaffiliation{Department of Chemistry, Norwegian University of Science and Technology, 7491 Trondheim, Norway}
\email{henrik.koch@sns.it}
\author{Stefano Corni}
\affiliation{Dipartimento di Scienze Chimiche and Padua Quantum Technologies Research Center, University of Padova, I-35131 Padova, Italy}
\alsoaffiliation{Istitute of Nanosciences, Consiglio Nazionale delle Ricerche CNR-Nano, I-41125 Modena, Italy}
\email{stefano.corni@unipd.it}


\begin{document} 
\singlespacing
\blfootnote{$^\Delta$ These authors contributed equally to the realization of the present work}
\maketitle

\begin{tocentry}
\begin{center}
\includegraphics[width=1\linewidth]{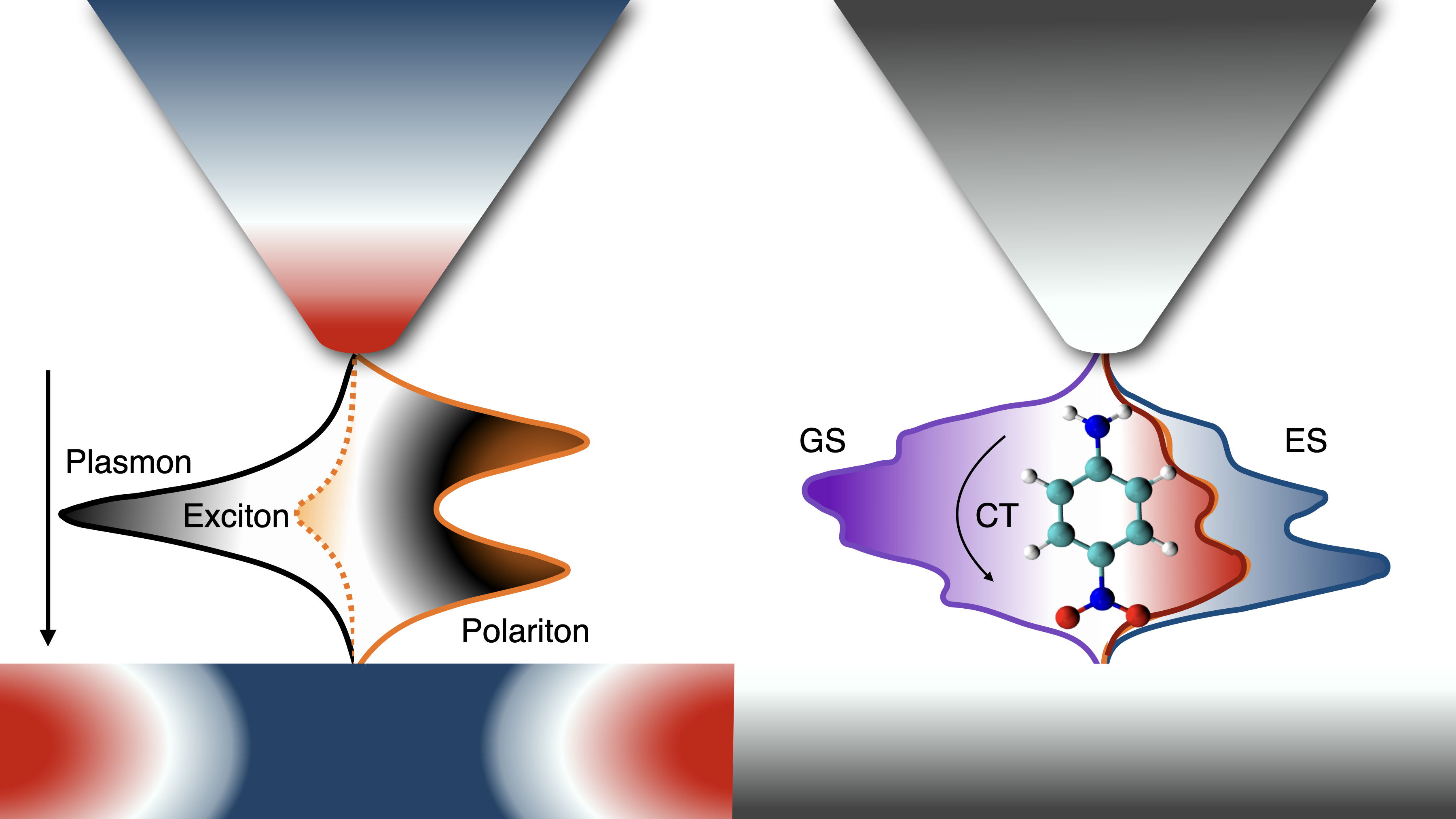}
\end{center}
\end{tocentry}

\begin{abstract}
Plasmonic nanocavities enable the confinement of molecules and electromagnetic fields within nanometric volumes. As a consequence, the molecules experience a remarkably strong interaction with the electromagnetic field, to such an extent that the quantum states of the system become hybrids between light and matter: polaritons. Here we present a non-perturbative method to simulate the emerging properties of such polaritons: it combines a high-level quantum chemical description of the molecule with a quantized description of the localized surface plasmons in the nanocavity. We apply the method to molecules of realistic complexity in a typical plasmonic nanocavity, featuring also a subnanometric asperity (picocavity). Our results disclose the effects of the  mutual polarization and correlation of plasmons and molecular excitations, disregarded so far. They also quantify to what extent the molecular charge density can be manipulated by nanocavities, and stand as benchmarks to guide the development of methods for molecular polaritonics.
\end{abstract}
\begin{spacing}{1.2}

Strong coupling between molecules and quantum plasmons\cite{qplas} in nanocavities\cite{nanocav1} leads to the formation of hybrid plasmon-molecule states: polaritons, or more specifically, plexcitons\cite{plexciton2}. These new states manifest distinct features compared to the original states\cite{pespec,musser:triplet,joel:inverting,climent:gs1}, potentially resulting in modified chemical/photochemical
reactivity\cite{ebbesen:vsc2,singlet_fission,fregoni:chem}
and relaxation dynamics\cite{mukamel:coin,jay:nonherm},
along with other coherent processes\cite{remi:ct,coccia:coherence2}. Modeling accurately the molecules, nanostructures and their coupling is of the utmost importance to support the experimental advances. Coupling descriptions proper for resonant optical cavities, such as using the molecular dipole only\cite{galego:cavity,mukamel:nadiab} should be amended\cite{neuman:dipole,aizpurua:tautom,slowik:quad} when the coupling affects the molecule on a sub-molecular level\cite{baumberg:qed,aizpurua:tautom,ters3}. 
 
There are currently no approaches that treat the coupled plasmon-molecule system non-perturbatively, \textit{i.e.} that include the relaxation of the molecular electronic density upon polariton formation together with plasmon-molecule correlation. In this work we extend the QED-CC method\cite{haugland2020qedcc,tor:inter} - already applied to resonant optical cavities - to realistic nanoplasmonic cavities. We quantize the plasmonic excitations starting from the classical dielectric description of nanostructures with arbitrary shapes. Each quantized plasmonic mode is associated with a surface charge density (discretized into point charges), analogous of a transition charge density in a molecular system. A similar quantization approach for plasmons in Drude metals, derived from macroscopic QED,\cite{Buhmann_2012,feist:macroqed,slowik:quad} was adopted in ref.\cite{neuman:dipole}. The quantization framework we are presenting accounts for the geometrical features of the nanoparticles setups without disregarding the molecular complexity. Indeed, coupled cluster (singles and doubles excitations) is recognized as gold standard in quantum chemistry\cite{Helgaker2000book}: we extend it to include self-consistent and correlated molecule-plasmon hybridization. 
With this method, we compute the interaction between a nanocavity realized after ref.\cite{aizpurua:tautom} and two realistic molecules: porphyrin and para-nitroaniline (PNA). For porphyrin, we highlight the role of the geometrical features of the system and the role of electron-plasmon self-consistency and correlation in polaritons, beyond the standard Jaynes-Cumming picture\cite{jaynes:cav}. We use PNA to quantify the same effects -- already predicted on the ground and excited states electron densities\cite{haugland2020qedcc} for a resonant cavity -- in the case of a nanocavity.

\section*{Results and Discussion}
\subsection*{Theory}
The first subsystem we focus on is the nanostructure.
We start from a classical dielectric perspective, namely computing the plasmon modes from a continuous medium model. As such, it is not an atomistic description and does not include dielectric non-local effects, electron spillout or chemical bonds. However, the description we adopt is surprisingly accurate\cite{antonio:classical}: it reproduces the inhomogeneous electric field due to atomistic irregularities in the nanostructures\cite{aizpurua:acsnano} even at sub-nanometric level\cite{aizpurua:nat}.
Our goal is to describe the nanostructure at quantum level as a material system, which differs from other approaches that quantize the electromagnetic field in the presence of dielectric media.\cite{Buhmann_2012,canquant}. 
By building on Polarizable Continuum Model applied to nanoparticles (PCM-NP) approach\cite{corni:natrev}, we shall achieve quantisation (Q-PCM-NP) in the quasi-electrostatic framework, where 
the retardation effects are disregarded.\cite{quasistatic}
Similar quantization frameworks for the plasmonic modes were used in different contexts not involving a molecule\cite{cherqui} or using a model description of the molecule\cite{truegler08} To classically compute the nanoparticle dielectric response properties, we solve Maxwell's equations for a continuous, frequency-dependent dielectric (nanostructure) under an external electromagnetic perturbation. In the quasi-static limit, the nanoparticle experiences an electrostatic potential at a given frequency, $\bm{V}(\omega)$, which induces polarisation charges $\bm{q}(\omega)$ on the nanoparticle surface. The equation defining such response charges is
\eqn{
	\bm{q}(\omega)=\bm{Q}^{IEF}(\omega)\bm{V}(\omega),
	\label{eq:asc}
}
where $\bm{Q}^{IEF}(\omega)$ is the frequency-dependent response to the external perturbation $\bm{V}(\omega)$.
We refer to IEF as to the specific Integral Equation Formalism\cite{tomasi:iefpcm1} formulation of the PCM problem.
In the present work, $\bm{V}(\omega)$ is the potential produced by the molecular transition density. The frequency-dependent linear charge density response function of the nanostructure is taken in a diagonal form\cite{pipolo:bem1}
\eqn{
\bm{Q}^{IEF}(\omega)=-\bm{S}^{-\frac{1}{2}}\bm{T}\bm{K}(\omega)\bm{T^\dagger}\bm{S}^{-\frac{1}{2}}
\label{eq:qief},
}
\eqn{
	K_{p}(\omega)=\frac{2\pi+\Lambda_{p}}{2\pi\frac{\varepsilon(\omega)+1}{\varepsilon(\omega)-1}+\Lambda_{p}}
\label{eq:kp}
}
where $\bm{K}$ is the diagonal linear-charge-response-matrix with eigenvalues $\Lambda_p$, $\varepsilon(\omega)$ is the frequency-dependent dielectric function and $\bm{S}$ is the matrix storing the electrostatic potential between discrete points of the dielectric. This is the starting point of the Q-PCM-NP quantization procedure.

\subsubsection*{Quantization of the plasmonic modes: Q-PCM-NP}
\begin{figure*}[ht!] 
	\centerline{\makebox[1.0\paperwidth][c]{
		\includegraphics[width=1\linewidth]{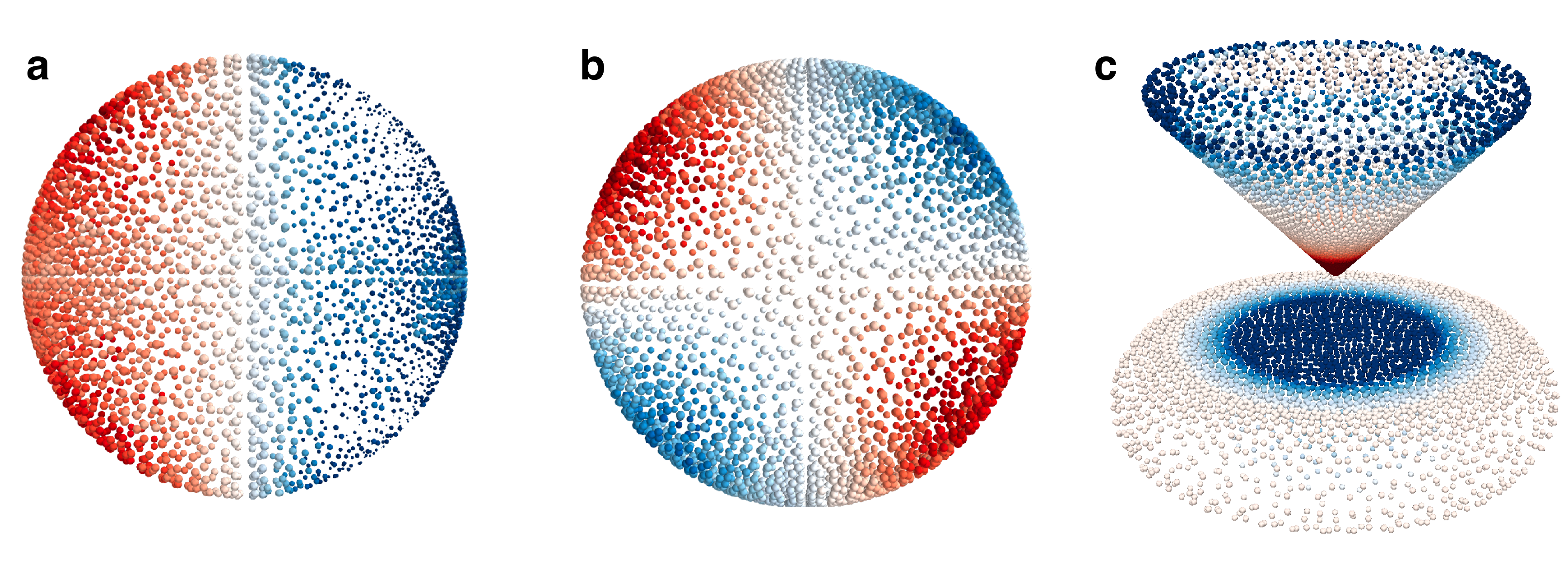}}}
	\caption{\textbf{Examples of plasmonic modes computed at Q-PCM-NP|} \textbf{a)}Example of plasmonic dipolar mode for a 10 nm sphere. The $p$ index associated to each set of charges does not take into account the $2l+1$ degeneracy for the modes of the sphere. Hence, the dipolar modes are $p=2$ to $p=4$. \textbf{b)} Example of a quadrupolar mode for the same sphere. Like for the dipolar case, the quadrupolar modes are $2l+1$ degenerate, hence they are from $p=5$ to $p=9$ and so on. \textbf{c)} Relevant plasmonic mode for a nanotip on a support. The mode is characterised by a strong charge density concentrated on the tip as in ref.\cite{aizpurua:tautom}. The dimensions of the system are about 10 by 10 nm (see Figure S1b)}.
\label{fig:modes}
\end{figure*}

We assume that the metal nanoparticle can be characterized by a Drude-Lorentz dielectric function
\eqn{
	\varepsilon(\omega)=1+\frac{\Omega_P^2}{\omega_0^2-\omega^2-i\gamma\omega},
\label{eq:drude}
}
where $\Omega_P^2$ is the squared plasma frequency of the bulk metal, $\omega_0$ is the natural frequency of the
bound oscillators (Lorentz model) and $\gamma$ is the damping rate.
By defining $\omega_p^2 = \omega_0^2+\left(1+\frac{\Lambda_p}{2\pi}\right)\frac{\Omega_P^2}{2}$ (neglecting the second-order term in $\gamma$), we retrieve (see Supporting Information) the full form of the response function $\bm{Q}^{IEF}(\omega)$ from eqs. \ref{eq:kp} and \ref{eq:drude}
\eqn{
	\bm{Q}^{IEF}_{kj}(\omega)=-\sum_p\left(\bm{S}^{-\frac{1}{2}}\bm{T}\right)_{k,p}\sqrt{\frac{\omega_p^2-\omega_0^2}{2\omega_p}}\left(\frac{1}{\omega_p+\omega+i\frac{\gamma}{2}}+\frac{1}{\omega_p-\omega-i\frac{\gamma}{2}}\right)\sqrt{\frac{\omega_p^2-\omega_0^2}{2\omega_p}}\left(\bm{T}^\dagger \bm{S}^{-\frac{1}{2}}\right)_{p,j}.
\label{eq:qief2}
}
Each matrix element of $\bm{Q}^{IEF}(\omega)$ is evaluated on representative points of the tesserae, $k$ and $j$. In eq. \ref{eq:qief2}, we recognize the spectral form of a linear response function\cite{boyd:nonlin}, $\bm{Q}^{quant}(\omega)$. Our goal now is to identify the quantities characterising the excited states of the nanostructure (i.e., plasmons) relevant to devise the coupling with the molecule.
For the quantum description of the nanostructure, the response function $\bm{Q}^{quant}(\omega)$ is formally written in terms of the surface charge operator $\hat q$\cite{ciro:jcp}.  We conveniently write $\bm{Q}^{quant}(\omega)$ matrix elements in its spectral representation\cite{jorg:damped} as 
\eqn{
	Q_{kj}^{quant}(\omega)=-\sum_p\left(\frac{\bra{0}\hat q_k\ket{p}\bra{p}\hat q_j \ket{0}}{\omega_p+\omega+i\gamma_p'}+ \frac{\bra{p}\hat q_k \ket{0}\bra{0}\hat q_j \ket{p}}{\omega_p-\omega-i\gamma_p'}\right),
	\label{eq:excresp}
}
By comparing eq. \ref{eq:qief2} and eq. \ref{eq:excresp},
we verify that $\omega_p$ represents the excitation frequencies of the plasmonic system, $\gamma_p'=\gamma/2$ for each plasmon state $p$ is a decay rate, while we can identify $\bra{0}\hat q_k\ket{p}$ with
\eqn{
	\bra{0}\hat q_k\ket{p}=\left(\bm{S}^{-\frac{1}{2}}\bm{T}\right)_{k,p}\sqrt{\frac{\omega_p^2-\omega_0^2}{2\omega_p}}=q_{p,k}.
	\label{eq:qp}
}
The label $q_{p,k}=\bra{0}\hat q_k\ket{p}$ represents the transition charge sitting on the $k$-th tessera associated to the mode $p$ and they are the analogous of transition densities in molecules. Expressed differently, the ensemble of the $q_{p,k}$ set of charges represents the normal mode of the plasmonic system at frequency $\omega_p$ (see Figures \ref{fig:modes}a, \ref{fig:modes}b and \ref{fig:modes}c for an example).
\noindent We are now in the position to write the quantum plasmonic Hamiltonian as
\eqn{
	\hat H_p=\sum_p\omega_p\left(\hat b_p^\dagger \hat b_p\right),
}
where $\omega_p$ is the $p$-mode frequency and $\hat b_p^\dagger$, $\hat b_p$ are the corresponding bosonic creation and annihilation operators.\cite{feist:macroqed,neuman:dipole} Assuming the  localized surface plasmons to be bosons is not a result of the present derivation (at the linear response level, a fermionic or bosonic excitation would yield the same $\bm{Q}^{quant}(\omega)$ ). It is rather based on the general behavior of plasmons in extended systems.\cite{qplas} In the Supporting Information, we report numerical checks of the present approach against analytical results for a point-dipole molecule interacting with a Drude metal nanosphere.\cite{delga:dipole} We can now write the general quantized plasmon-molecule Hamiltonian as
\eqn{
	\hat H= \hat H_{mol} + \hat H_{p}+ \hat H_{int},
	\label{eq:hamiltonian}
}
where $\hat H_{mol}$ is the standard molecular Hamiltonian (which we reduce to the electronic Hamiltonian). Making use of the response charges for the quantized plasmon, the interaction term $\hat H_{int}$ reads\cite{ciro:jcp}
\eqn{
	\hat H_{int}=\sum_{p,j} q_{p,j}\hat V_j \left(\hat b_p^\dagger+\hat b_p\right),\label{eq:hint}
}
where $\hat V_j$ is the molecular electrostatic potential operator, evaluated at each tessera representative point $j$. 
Our formulation of $\hat H_{int}$ presents two major advances with respect to simplified models (see Supporting Information for the formal comparison): we obtain an intuitive representation of the inhomogeneous electromagnetic field in terms of polarization charges and the interaction is general on the molecular side, meaning that it can be interfaced to advanced electronic structure models. Through this interface, our method can describe phenomena occurring in both the weak and strong coupling regime.\cite{flick:weaktostrong,schaefer:qed}
Radiative corrections to the quasi-static description, related to radiation reaction,\cite{novotny:book} can also be included in this model, at the quantum level. The plasmon radiative decay rate can be calculated from the transition dipole moment associated to the charges $q_{p,j}$ and it can then be added to the non-radiative lifetime $\gamma_p$. For the systems treated in the numerical section, this correction is negligible. An estimate of the molecule-plasmon coupling $g_{n}$ between the plasmonic mode $p$ and the molecular electronic transition from the ground state $S_0$ to the excited state $S_n$ is given by
\eqn{
	g_{n}=\expect{S_0,1}{\hat H_{int}}{S_n,0}=\sum_j q_{p,j}V^{(0,n)}_j,\label{eq:g_pmn}
}
Here $0$ and $1$ are the occupation numbers of the plasmonic mode $p$. The coupling $g_n$ is the central quantity in simplified approaches to strong coupling, such as the Jaynes-Cummings model (JC).

\subsubsection*{Extending QED coupled cluster (QED-CC) to plasmon-molecule systems}
Determining the eigenfunctions of the plasmon-molecule Hamiltonian in eq. \ref{eq:hamiltonian} requires an accurate description of the molecule. In this section, we outline the extension of the QED-CC treatment (already applied to resonant cavities\cite{haugland2020qedcc,tor:inter}) to the case of plasmons. The detailed algorithm is available in the QED-CC dedicated section of the Supporting Information. The simplest description of the molecule without plasmon interactions is a single Slater determinant, $\ket{\text{HF}}$ (Hartree-Fock). To include electron correlation
we rely on coupled cluster theory\cite{Helgaker2000book}. More explicitly, we apply the exponential of the cluster operator to the $\ket{\text{HF}}$ determinant,
\begin{equation}
    \ket{\text{CC}} = e^{\hat T} \ket{\text{HF}}.
\end{equation}
The cluster operator $\hat T$ is expressed as
\begin{equation} \label{eq:cluster_e}
    \hat T = \hat T_1 + \hat T_2 + \dots + \hat T_{N_e}
\end{equation}
where $T_1$ are linear combinations of single excitations, $T_2$ are linear combinations of double excitations and so on. Inspired by the formal similarities of the interaction in molecule-plasmon and molecule-photon systems, we extend the newly developed quantum electrodynamics coupled cluster theory (QED-CC)\cite{haugland2020qedcc} from photons to plasmons. Here the plasmon-molecule interaction is described non-perturbatively, allowing also the ground state to couple to the electromagnetic field.\cite{tor:inter}. In QED-CC, the cluster operator $\hat T$ also includes the bosonic creation and annihilation operators $\hat b^\dagger_p$, $\hat b_p$ of the quantised plasmon mode $p$: 
\begin{align} 
    \hat T &= \hat T_1 + \hat T_2 + \hat S^1_1 + \hat S^1_2 + \hat \Gamma^1.
\end{align}
Here we identify three contributions: electronic ($\hat T_1$, $\hat T_2$), plasmonic ($\hat \Gamma^1$) and mixed plasmon-molecule excitations ($\hat S_1^1$, $\hat S_2^1$). By the effect of such excitations, QED-CCSD-1 (SD denotes single and double excitation of the electronic subspace, whereas 1 is the excitation order of the plasmonic modes) includes the correlation 
induced on the electronic states by the plasmonic transition. We note that, in the limit where the interaction $\hat g \rightarrow 0$, CCSD and QED-CCSD-1 are equivalent.

\subsection*{Calculations on realistic molecules}
\subsubsection*{Free-Base Porphyrin}

\begin{figure}[ht!]
    \centering
    \includegraphics[width=\textwidth]{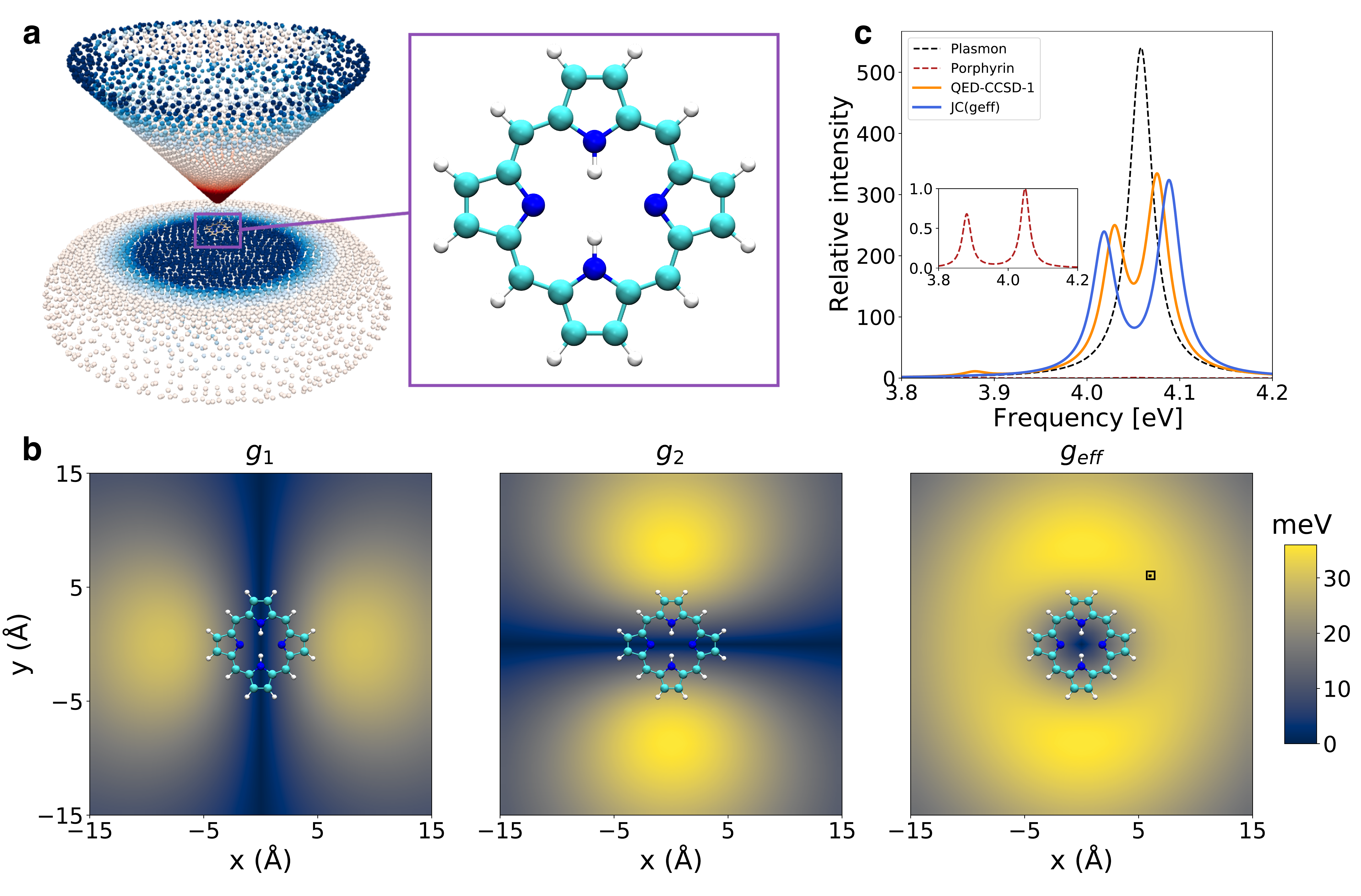}
    \caption{\textbf{Interaction between the free-base porphyrin and a nanotip|} \textbf{a)} Setup for porphyrin interacting with a plasmonic nanotip used in Panels b and c, where the nanotip has transition energy $\omega=4.06$ eV.
    \textbf{b)} Map of the plasmon-molecule coupling $g$ for two excitations in porphyrin (4.05 eV, 4.07 eV) and the effective coupling $g_{\text{eff}}=\sqrt{g_1^2+g_2^2}$. Axis $x$ and $y$ indicate the displacement of the nanotip from the centre of the molecule.
    \textbf{c)} Oscillator strengths (intensity) for porphyrin displaced 6 {\AA} along $x$ and 6 {\AA} along $y$ (see black square in b)). Intensities are relative to porphyrin's strongest transition. The inset shows the same plot with only porphyrin. The CCSD Rabi splitting is calculated using a two-state Jaynes-Cummings model (JC) with intensity relative to $g_{\mathrm{eff}}$.}
    \label{fig:porphyrin_fig}
\end{figure}

In this section, we exploit Q-PCM-NP coupled to QED-CCSD-1 and EOM-CCSD to simulate plasmon-molecule interactions. Here we consider a free-base porphyrin coupled to the plasmonic nanotip, as sketched in Figure \ref{fig:porphyrin_fig}a. A similar setup was adopted in refs.\cite{neuman:dipole,aizpurua:tautom}
According to eq.\ref{eq:g_pmn}, we define the couplings between the ground state $\ket{S_0,1}$ of the free-base porphyrin and the $\ket{S_1,0}$, $\ket{S_2,0}$ states as $g_1$ and $g_2$ respectively. In Figure \ref{fig:porphyrin_fig}b, we show the associated interaction maps, where the couplings $g_1$ and $g_2$ are displayed as a function of the molecule-nanotip displacement. The two quasi-degenerate dipolar transitions of free-base porphyrin at $4.05$ eV ($S_1$) and $4.07$ eV ($S_2$) have different symmetries. Depending on the tip displacement, the interaction element $g$ varies between 0 meV at the centre to 36 meV at 6 {\AA} displacement. 
By displacing the tip accordingly, a polariton is created with either of the two states.
When combining the two quasi-degenerate transitions, $g_{\text{eff}}=\sqrt{g_1^2 + g_2^2}$, a doughnut-like shape of the transition density is obtained, in agreement with previous results on zinc phtalocyanine\cite{neuman:dipole}.
We stress that this doughnut-like shape is not observable without including the geometry of the nanostructure.
In Figure \ref{fig:porphyrin_fig}c, we show the absorption spectrum of the relaxed tip-molecule complex (QED-CCSD-1) and compare it with
the polaritons obtained by the Jaynes-Cummings model (see the Supporting Information for details).
The point at which we perform the calculations is highlighted as the black square in panel $g_{\mathrm{eff}}$ of Figure \ref{fig:porphyrin_fig}b. 
In QED-CCSD-1, the light-matter system is allowed to correlate both the ground and the excited states, guaranteeing accurate results with respect to the truncated basis. Even more, the interactions between all singly and doubly excited determinants, with and without plasmons, are included. Conversely, the JC case is treated by correlating the electronic states only, hence neglecting the mixed molecule-plasmon correlation contributions. As a further difference between JC and QED-CCSD-1, the interaction in the former is a simple norm of the $g_1$ and $g_2$ interactions and it does not allow for the $S_1$ and $S_2$ states to interact through the plasmon. 
For QED-CCSD-1, the presence of the polaritonic correlation and several electronic states reveals that the plasmon contribution is distributed over other electronic states displayed as the small peak at 3.88 eV in Figure \ref{fig:porphyrin_fig}c. At the same time, the Rabi splitting between the two main polaritonic peaks is reduced (46 meV) with respect to JC (70 meV). While the small peak is due to the inclusion of different states, the difference in the Rabi splitting is due to the inclusion of the mixed plasmon-molecule correlation. To corroborate this view, we compare in the Supporting Information an extended JC model using three states with the QED-CCSD-1. 
Although the description improves with the extended JC model, it still misses the intensities and the positions of the polaritonic peaks, confirming the role of the mixed plasmon-molecule correlation in the description.

\subsubsection*{Para-nitroaniline}
\begin{figure}[h!]
    \centering
    \includegraphics[width=1\linewidth]{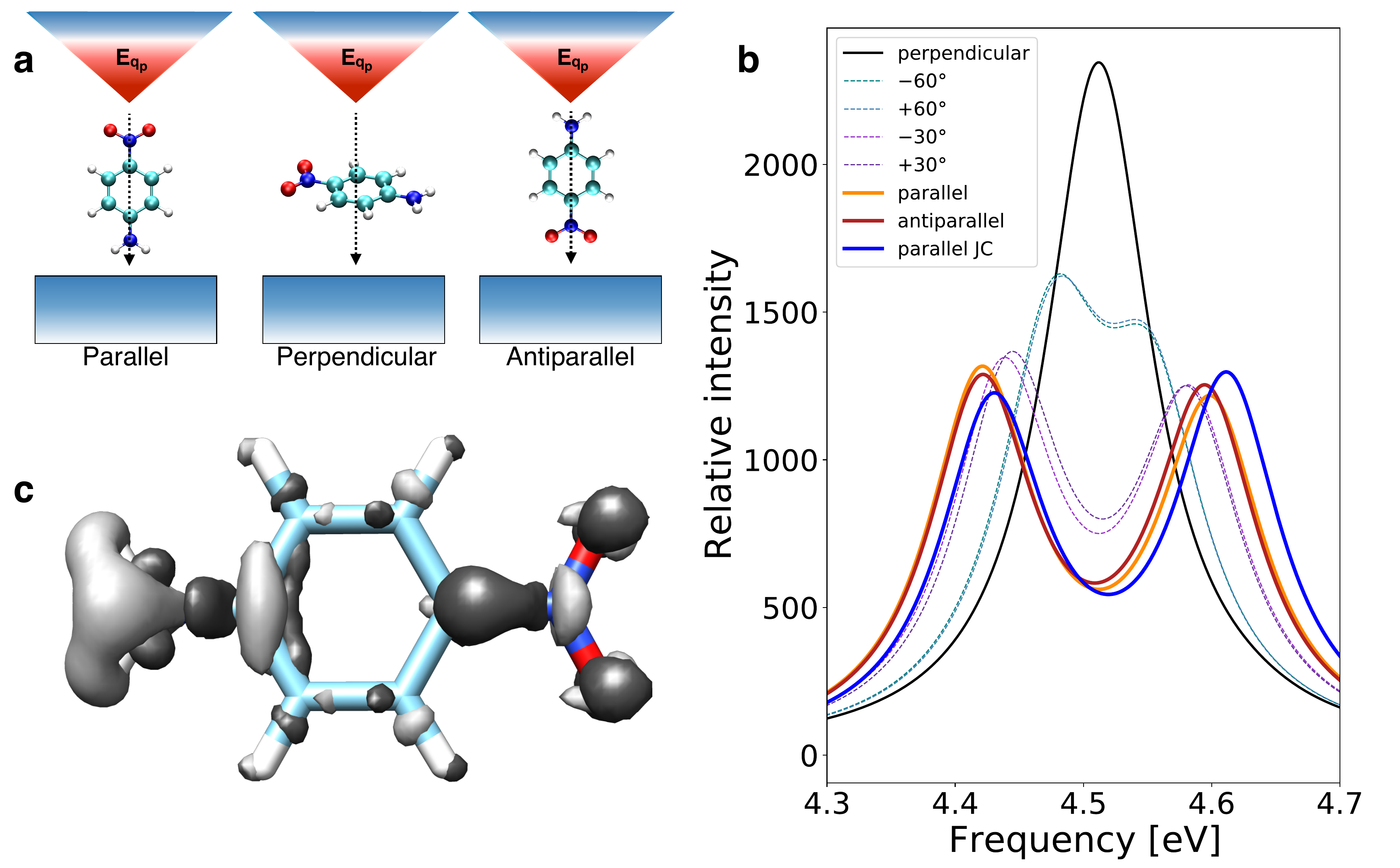}
    \caption{\textbf{Quantum coupling between a PNA molecule and a nanotip|} \textbf{a)} Sketch of the molecule interacting with the plasmonic nanotip. The parallel orientations identify the maximum plasmon-molecule coupling condition, whereas the perpendicular one corresponds to the minimum coupling.
    \textbf{b)} PNA oscillator strength (intensity) at different orientations inside a plasmonic nanotip-cavity. Each calculation is performed at different molecular orientations around the nuclear center of charge. The plasmon is resonant with the bright molecular transition (4.52 eV). The comparison between the realistic structure of the molecule (QED-CCSD) and the 2-levels Jaynes Cummings model (JC) is shown by the full colored lines.
    \textbf{c)} Plasmon-induced difference in PNA's ground-state density, when PNA is parallel with the plasmonic mode of frequency $\omega=4.52$ eV. Black and white isosurfaces indicate increased and reduced density ($\pm10^{-3}\, e^{-}/{a_0^3}$).}
    \label{fig:pna}
\end{figure}
Para-nitroaniline (PNA) presents an intense absorption peak in the UV range. We hence expect large Rabi splittings when the PNA is appropriately aligned with the electromagnetic field of the nanotip. The transition density associated to the bright transition of PNA at $\sim$4.5 eV is parallel to the long axis of the molecule, making it an excellent candidate for single-molecule strong coupling. 

\begin{figure}[h!]
    \centering
    \includegraphics[width=0.9\textwidth]{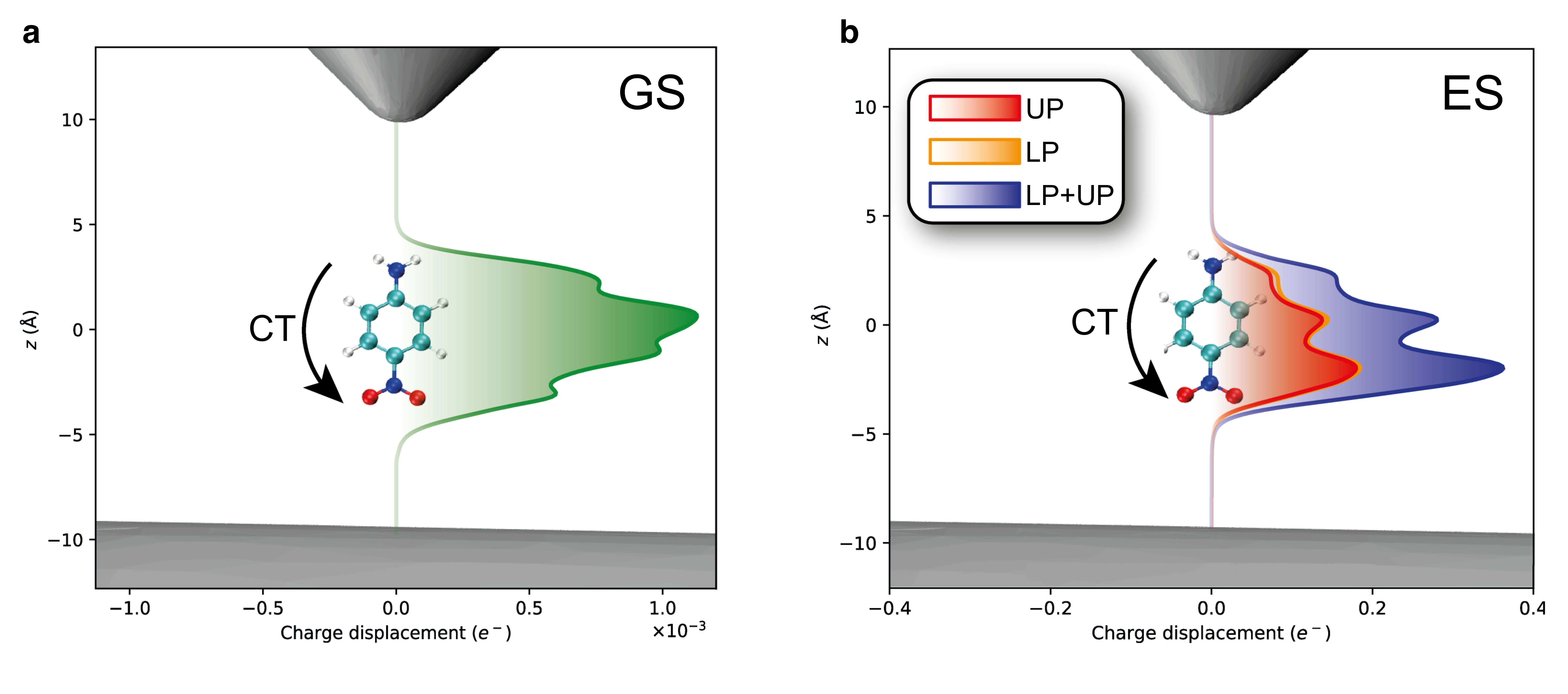}
    \caption{\textbf{Charge displacement analysis of PNA.} The molecule is placed parallel with respect to the plasmonic nanotip. The direction of charge transfer (CT) is along the arrow. \textbf{a)} Plasmon-induced density difference in the ground state. \textbf{b)} Charge transfer excitation around 4.5 eV. The red and orange areas are upper (UP) and lower (LP) polaritons respectively. The blue curve is the sum of red and orange.}
    \label{fig:pna_cdz}
\end{figure}
In Figure \ref{fig:pna}, we investigate the coupling of the PNA molecule with the nanotip. We identify the maximum (parallel) and the minimum (perpendicular) coupling conditions, as sketched in Figure \ref{fig:pna}a and plot the Rabi splitting of the plasmon-molecule coupled system in Figure \ref{fig:pna}b. 
When the molecular transition dipole is perpendicular to the plasmon mode, there is little-to-no effect on the spectrum.
Instead, when the molecule is gradually rotated to be parallel to the mode, the Rabi splitting increases until the maximum of 179 meV (parallel). By gradually rotating the molecule in the opposite direction (anti-parallel) we observe a slightly smaller splitting (175 meV). The 4 meV difference between the parallel and anti-parallel orientations is a signature of the realistic molecular description. Indeed, a dipolar interaction would cause result in a symmetric spectrum, with a Rabi splitting independent of the parallel/anti-parallel orientation. In addition, we observe a red-shift of the polaritonic peaks when including the mixed plasmon-molecule correlation energy via QED-CCSD-1.\\

The Jaynes-Cummings model fails at describing ground state properties, as no change in the molecular ground state is predicted by the model. By instead using QED-CCSD-1, we show in Figure \ref{fig:pna}c how the ground-state electronic density changes due to the molecule-plasmon interaction. Since the coupling in the present case is smaller than the one considered in ref.\cite{haugland2020qedcc}, the charge localization effects as discussed in the same reference are reduced. We quantify this effect
in Figure \ref{fig:pna_cdz}a. 
We note the opposite trend to what was described in ref.\cite{haugland2020qedcc}: here, the plasmon induces a small charge transfer from the donor to the acceptor. This inversion is a signature that the molecules interact differently with plasmonic nanotips compared to optical cavities, whereas the Jaynes-Cummings model treats them equivalently. The charge transfer properties of the polariton states are also substantially changed, reminiscent of what is shown in ref. \cite{haugland2020qedcc}. However, in the present case, the sum of the polaritons' charge transfer adds up to the charge transfer of the singlet state without the plasmons, as displayed in Figure \ref{fig:pna_cdz}b. This again highlights that interactions with nanoparticle plasmons and optical cavities are not equivalent.
The plasmon-induced changes in the electronic density also imply a modification of other ground-state properties as well as the potential energy surfaces. One example is the minor induced change of the dipole moment ($-0.03$ D). One important limitation in this ground-state study is the inclusion of only a single plasmon mode: higher energy modes are expected to contribute to the ground-state coupling and enhance the minor effects seen here, as the ground-state correlation is not a resonant property\cite{tor:inter}.

\section*{4. Conclusions}
In this work we have introduced a new methodology, based on PCM-NP\cite{corni:response,mennucci:pcm} and QED-CC\cite{haugland2020qedcc}, to treat the quantum coupling
between nanoparticles of arbitrary shapes and molecules.
We compared our description of polaritons in free-base porphyrin with the simplified Jaynes-Cummings model, highlighting the role of the mixed molecule-plasmon correlation. Finally, we have analyzed the effects on the ground state of PNA in a realistic plasmonic nanocavity setup, showing that the modification of ground states density previously investigated by QED-CC is still present. Our representation of quantum plasmonic modes as apparent surface charges, in synergy with coupled cluster, provides a simple-yet-accurate interface to tackle mixed nanoparticle-molecules systems. 
In conclusion, we believe our method 
provides solid ground to support and guide experiments investigating and manipulating the chemical properties at the sub-molecular level. Accurate calculations of plexciton-affected reaction-barriers are already possible, as well as exploring effects related to chiral nanotips. The method is also prone to various extensions on the nanostructure side: the inclusion of multiple EM modes or the extension to classical atomistic models\cite{Jensen2009,jensen:atom2, Giovannini2019}. Finally, the model may assess the properties of molecules within plasmonic nanocavities in turn hosted by optical cavities, a setup whose experimental exploration has been recently started.\cite{Baranov2020}\\

Supporting Information detail the theoretical and computational specifics for Q-PCM-NP and QED-CC, and provide further benchmarks and supplementary results. The computational details for the reproducibility of the calculations and the codes availability are reported in the dedicated section of the Supporting Information. The material is available free of charge at http://pubs.acs.org.

\begin{acknowledgement}
 We acknowledge Enrico Ronca's contributions to the charge displacement analysis. The authors also thank Antonio I. Fernandez Dominguez and Johannes Feist for the fruitful scientific discussion on the field quantization. This work has been funded by the European Research Council through grants ERC-2015-CoG-681285 (J. Fregoni, PI Stefano Corni), and the Research Council of Norway through FRINATEK projects 263110 and 275506 (T. S. Haugland, PI Henrik Koch). We acknowledge computing resources through UNINETT Sigma2 - the National Infrastructure for High Performance Computing and Data Storage in Norway, through project number NN2962k. We further acknowledge computer resources from the HPC Center at Scuola Normale Superiore di Pisa.
\end{acknowledgement}

\end{spacing}

\singlespacing
\providecommand{\latin}[1]{#1}
\makeatletter
\providecommand{\doi}
  {\begingroup\let\do\@makeother\dospecials
  \catcode`\{=1 \catcode`\}=2 \doi@aux}
\providecommand{\doi@aux}[1]{\endgroup\texttt{#1}}
\makeatother
\providecommand*\mcitethebibliography{\thebibliography}
\csname @ifundefined\endcsname{endmcitethebibliography}
  {\let\endmcitethebibliography\endthebibliography}{}

\end{document}


\blfootnote{$^*$ These authors contributed equally to the realization of the present work}
\maketitle
\begin{spacing}{1.2}

\clearpage
\section*{Q-PCM-NP}
\subsection*{Response of the plasmonic nanostructures to an electromagnetic field: the PCM-NP equations in the diagonal formulation}
The first step is to choose a convenient description of the classical dielectric response of the nanoparticle. Our approach builds on the PCM-NP formalism\cite{corni:response,pipolo:bem1}. In particular, we make use of PCM relations from the integral equation formalism\cite{tomasi:iefpcm1}.\\

\begin{figure*}[ht!] 
	\centerline{\makebox[1.0\paperwidth][c]{
		\includegraphics[width=1\linewidth]{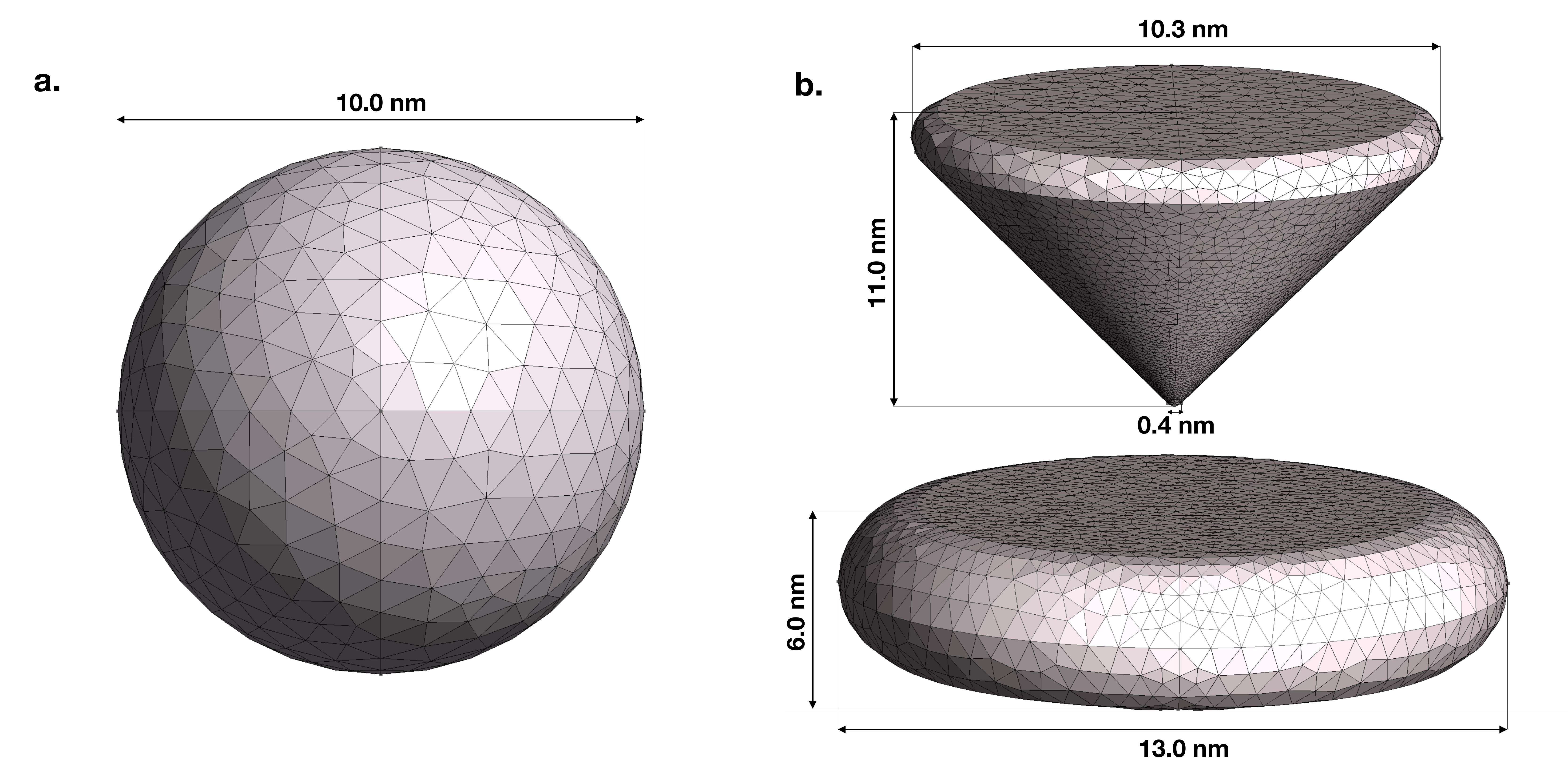}}}
	\caption{\textbf{Nanoparticles surfaces discretised with BEM|} \textbf{a)} Surface discretization of a 10 nm spherical nanoparticle, which will be used in the tests of the presented method. \textbf{b)} Discretisation of a nanotip setup. The dimensions are reproduced after ref.~\cite{aizpurua:tautom}.}\label{fig:bem}
\end{figure*}

\noindent To classically compute the nanoparticle surface properties means to classically solve Maxwell's equations for a continuous, frequency-dependent dielectric (nanostructure) under an external perturbation. Put differently, simulating the plasmons of a nanoparticle calls for the computation of the nanoparticle's linear response to the external perturbation. In the quasi-static limit, the nanoparticle experiences an electrostatic potential $\bm{V}(\omega)$ acting on the nanoparticle surface, where $\omega$ is the frequency associated to the external potential (\textit{e.g.} produced by a molecular transition). Such potential induces polarisation charges. A commonly adopted solution to the electrostatic problem is to resort to discretisation techniques (Figures~\ref{fig:bem}a and \ref{fig:bem}b) such as the boundary element method (BEM)\cite{Fuchs1975,abajo,corni:response,GarciadeAbajo2002,ulrich:mnpbem1}.\\

\noindent Within such formulation, a set of apparent surface charges (ASC) sitting on the
representative points of the nanoparticle discretised surface is used to represent the electrostatic potential that solves the Poisson equation. We can consider the ASC as the representation of the linear dielectric response of the nanoparticle to the external perturbation at different $\omega$. 
The IEF-BEM equation defining the ASC ($\bm{q}$) is
\eqn{
	\bm{q}(\omega)=\bm{Q}^{IEF}(\omega)\bm{V}(\omega)
	\label{eq:asc}
}
where $\bm{Q}^{IEF}(\omega)$ is the frequency-dependent response matrix to the external perturbation $\bm{V}(\omega)$. It describes the redistribution of the surface charges in the NP when the system is subjected to an external perturbation which, here, will be the electric field associated to the molecular transition density. $\bm{Q}^{IEF}(\omega)$ is computed following the standard IEF formalism as\cite{pipolo:bem2}:
\eqn{
	\bm{Q}^{IEF}(\omega)=-\bm{S^{-1}}\left(2\pi \frac{\varepsilon(\omega)+1}{\varepsilon(\omega)-1}\bm{I}+\bm{DA} \right)^{-1}\left(2\pi\bm{I}+\bm{DA}\right).
\label{eq:resp}}
Here, $\varepsilon(\omega)$ is the dielectric function of the medium, $\bm{A}$ is a diagonal matrix containing the area of each discrete surface element (tessera), while $\bm{S}$ and $\bm{D}$ are the matrix representations of Calderon's projectors, respectively storing information on the electrostatic potentials and the electric field between charges sitting on different tesserae:
\eqn{
 S_{ij}=\frac{1}{\abs{\bm{s_i}-\bm{s_j}}}\qquad D_{ij}=\frac{\left(\bm{s_i}-\bm{s_j}\right)\cdot \bm{n_j}}{\abs{\bm{s_i}-\bm{s_j}}^3},
	\label{eq:sd}
}
for the out of diagonal elements (for diagonal elements see, e.g., ref. \cite{DallOsto2020}) where $\bm{s_i}$ and $\bm{s_j}$ collect the coordinates of the $i-$th and $j-$th tessera and $\bm{n_j}$ is the unit vector directed along the normal of the $j-$th tessera. Examining the response function in eq. \ref{eq:resp}, the frequency dependence is embodied in the choice of the dielectric function $\varepsilon(\omega)$. Following the procedure outlined in previous works \cite{pipolo:bem1}, it is possible to reformulate the matrix in a diagonal form:
\eqn{
	\bm{Q}^{IEF}(\omega)=-\bm{S}^{-\frac{1}{2}}\bm{T}\left(2\pi \frac{\varepsilon(\omega)+1}{\varepsilon(\omega)-1}\bm{I}+\bm{\Lambda} \right)^{-1}\left(2\pi\bm{I}+\bm{\Lambda}\right)\bm{T}^\dagger\bm{S}^{-\frac{1}{2}},
\label{eq:bemdiag}
}
Here, the diagonal matrix $\bm{\Lambda}$ and the matrix $\bm{T}$ collect the eigenvalues and eigenvectors of $\bm{S}^{-\frac{1}{2}}\bm{DA}\bm{S}^{\frac{1}{2}}$, which is (formally) symmetric. A compact form to indicate the diagonalized version of $\bm{Q}^{IEF}(\omega)$ is obtained when collecting all the central terms of eq. \ref{eq:bemdiag} into a diagonal matrix $\bm{K}(\omega)$:
\eqn{
\bm{Q}^{IEF}(\omega)=-\bm{S}^{-\frac{1}{2}}\bm{T}\bm{K}(\omega)\bm{T^\dagger}\bm{S}^{-\frac{1}{2}}
\label{eq:qief},
}
\eqn{
	K_{p}(\omega)=\frac{2\pi+\Lambda_{p}}{2\pi\frac{\varepsilon(\omega)+1}{\varepsilon(\omega)-1}+\Lambda_{p}}.
\label{eq:kp}
}
$K_{p}$ and $\Lambda_p$ are the respective diagonal elements of the $\bm{K}(\omega)$ and $\bm{\Lambda}$ matrices.
By substituting in the expression above the Drude-Lorentz dielectric function $\varepsilon(\omega)$:
\eqn{
	\varepsilon(\omega)=1+\frac{\Omega_P^2}{\omega_0^2-\omega^2-i\gamma\omega},
\label{eq:drude}
}
we make the frequency dependence of $K_p$ explicit:
\eqn{
	K_p(\omega)=\frac{\left(1+\frac{\Lambda_p}{2}\right)\frac{\Omega_P^2}{2}}{\omega_0^2-\omega^2-i\gamma\omega+\left(1+\frac{\Lambda_p}{2\pi}\right)\frac{\Omega_P^2}{2}}.
}
We define the plasmon frequency as $\omega_p^2 = \omega_0^2+\left(1+\frac{\Lambda_p}{2\pi}\right)\frac{\Omega_P^2}{2}$ (where we have neglected a second-order term in $\gamma$) and obtain
\begin{align}
	K_p(\omega)=&\frac{\omega_p^2-\omega_0^2}{\omega_p}\frac{\omega_p}{\left(\omega_p-\omega-i\frac{\gamma}{2}\right)\left(\omega_p+\omega+i\frac{\gamma}{2}\right)}\\
=&\frac{\omega^2_p-\omega^2_0}{2\omega_p}\left(\frac{1}{\left(\omega_p+\omega+i\frac{\gamma}{2}\right)}+\frac{1}{\left(\omega_p-\omega-i\frac{\gamma}{2}\right)}\right).\label{eq:response}
\end{align}
By making use of this form for $K_p(\omega)$ and of eq. \ref{eq:qief}, we retrieve the full form of the response function $\bm{Q}^{IEF}(\omega)$ as:
\eqn{
	\bm{Q}^{IEF}_{kj}(\omega)=-\sum_p\left(\bm{S}^{-\frac{1}{2}}\bm{T}\right)_{k,p}\sqrt{\frac{\omega_p^2-\omega_0^2}{2\omega_p}}\left(\frac{1}{\omega_p+\omega+i\frac{\gamma}{2}}+\frac{1}{\omega_p-\omega-i\frac{\gamma}{2}}\right)\sqrt{\frac{\omega_p^2-\omega_0^2}{2\omega_p}}\left(\bm{T}^\dagger \bm{S}^{-\frac{1}{2}}\right)_{p,j}.
\label{eq:qief2}
}

\subsection*{Correspondence between macroscopic QED charges and Q-PCM-NP: the Drude dielectric limit}

In this section, we compare the Drude limit ($\omega_0=0$) of the Q-PCM-NP formulation to the macroscopic QED charges. To do so, we report eq. 8 from the main text, representing the set of response charges associated to a plasmonic mode $p$:

\eqn{
	\bra{0}\hat q_k\ket{p}=\left(\bm{S}^{-\frac{1}{2}}\bm{T}\right)_{k,p}\sqrt{\frac{\omega_p^2-\omega_0^2}{2\omega_p}}=q_{p,k}.
	\label{eq:qp}
}

\noindent The individual sets of charges describing the quantised plasmons with the Q-PCM-NP method are analogous to the ones obatined in ref.\cite{neuman:dipole}, based on macroscopic QED for a Drude dielectric.
There, the charges are defined via the right eigenvectors $\sigma_p(\mb{s})$ of an integral operator kernel $\mathcal{F}(\bm{s},\bm{s}')$ that, once integral equations are discretized, is equivalent to $\mb{D}^\dagger$ in our notation. More specifically
\eqn{
\int_{\partial V_{pl}} \mathcal{F}(\mb{s},\mb{s}')\sigma_p(\mb{s}')~d^2\mb{s}'=\lambda_p \sigma_p(\mb{s})
}
with (in our notation and using atomic units):
\eqn{\int_{\partial V_{pl}} \int_{\partial V_{pl}}  \frac{\sigma_p^*(\mb{s}') \sigma_p(\mb{s})}{|\mb{s}-\mb{s}'|} d^2\mb{s} \, d^2\mb{s}'= \frac{1}{2} \sqrt{\frac{\Omega_P^2}{2} (1+\frac{\lambda_p}{2\pi})}
}
where ${\partial V_{pl}}$ is the surface of the nanoparticle. In the discretized version and the present paper notation, such equations read:
\eqn{
\mb{AD}^\dagger \mb{q}_p=\lambda_p \mb{q}_p
\label{eq:ad}
}
\eqn{\mb{q}_p^\dagger\mb{S}\mb{q}_p=\frac{1}{2} \sqrt{\frac{\Omega_P^2}{2} (1+\frac{\lambda_p}{2\pi})}}

By multiplying eq.(\ref{eq:ad}) on the left by $\mb{S}^{1/2}$, and using $\mb{S}^{1/2}\mb{AD}^\dagger S^{-1/2}=\mb{S}^{-1/2}\mb{DA}\mb{S}^{1/2}$ we get:
\eqn{
\mb{S}^{-1/2}\mb{DA}\mb{S}^{1/2} \mb{S}^{1/2}\mb{q}_p=\lambda_p \mb{S}^{1/2}\mb{q}_p
\label{eq:sad}
}
$\Lambda_p$ and the $p$ column of the matrix $\mb{T}$ in the main text are by construction the eigenvalue and the corresponding eigenvector of the matrix $\mb{S}^{-1/2}\mb{DA}\mb{S}^{1/2}$\cite{pipolo:bem1}. Therefore, $\lambda_p=\Lambda_p$,  $\mb{q}_p^\dagger\mb{S}\mb{q}_p=\frac{1}{2}\omega_p$ for Drude model ($\omega_0=0$) and $\alpha_p \mb{S}^{1/2}\mb{q}_p=\mb{T}_p$, where $\alpha_p$ is a normalization factor to be determined, that can be taken  real. To obtain the latter, we note that:
\eqn{1=\mb{T}_p^\dagger \mb{T}_p=\alpha_p^2\mb{q}_p^\dagger\mb{S}\mb{q}_p=\alpha_p^2 \frac{1}{2} \omega_p }
Therefore, $\alpha_p=\sqrt{2/\omega_p}$ and $\sqrt{2/\omega_p} \mb{S}^{1/2}\mb{q}_p=\mb{T}_p$, that is:
\eqn{\mb{q}_p=\sqrt{\frac{\omega_p}{2}}\mb{S}^{-1/2}\mb{T}_p}
that is the expression given in eq. \ref{eq:g_p} of the main text, once it is reduced to the Drude model ($\omega_0=0$). Therefore, our result does match the $\mb{q}_p$ from ref.\cite{neuman:dipole} in the case of a Drude metal.

\subsection*{Simplified coupling models}

In the main text we described how fully correlated plasmon-molecules wave functions can be obtained with QED-CCSD-1. In this section we shall derive approximate expressions of the plasmon-molecule coupling strength based on the properties of the unperturbed gas-phase molecule, and then by further reducing it to a point dipole. The former is an ubiquitous assumption in the theoretical descriptions of molecule-plasmonic nanostructrure strong coupling, whose consequences we are here in the position to check for the first time. The latter is a common assumption, whose general validity has been already questioned in the field of strong coupling, \cite{neuman:dipole} and before in the field of surface enhanced spectroscopies.\cite{corni:response,Corni2001cpl}\\

\noindent In eq. 9 of the main text, we show the plasmon-molecule interaction in the quasi-static limit. Based on such equation, we define - for each plasmonic mode $p$ - the plasmon-molecule coupling operator $\hat g_p$ as
\eqn{
	\hat g_p=\sum_j q_{p,j}\hat V_j \label{eq:g_p},
}
hence reducing the interaction Hamiltonian to:
\eqn{
	\hat H_{int}=\sum_p \hat g_p \left(\hat b_p^\dagger+\hat b_p\right).
}

\noindent The simplest approximation to the plasmon-molecule coupling strength $g_p$ for a specific plasmon mode $p$ is defined by the matrix elements of $\hat H_{int}$ on the minimal basis of non-interacting states required to describe the polariton formation,  i.e., $\ket{S_0}\otimes\ket{1}=\ket{S_0,1}$ and $\ket{S_1}\otimes\ket{0}=\ket{S_1,0}$.  $S_0$ and $S_1$ are the ground and the first excited state of the molecule, respectively; $\ket{0}$ and $\ket{1}$ are the plasmonic states characterized by the occupation number of the plasmonic mode we are focusing on (either 0 or 1 in this example). Using eq. \ref{eq:g_p}, one gets:
\eqn{
	g^{bem,full}_p=\expect{S_0,1}{\hat H_{int}}{S_1,0}=\expect{S_0,1}{\hat g_p \hat b^\dagger}{S_1,0}=\sum_j q_{p,j}V^{(S_0,S_1)}_j,
}
where $V^{(S_0,S_1)}_j$ is the potential originated by the $S_0\rightarrow S_1$ transition on the $j$-th tessera. The superscript $bem,full$ reminds us that this is an expression obtained by the {\it BEM} formulation with {\it full} consideration of the spatial extent of the molecule, i.e., no approximation of the molecule as a point dipole. To find an expression for $V^{(S_0,S_1)}_j$, it is convenient to describe the local behaviour of the molecule in terms of one-particle operators in second quantization. The potential acting on the $j$-th tessera is then the potential associated to the one-particle electron density operator $\hat \rho(\bm{r})$, written in terms of the molecular orbitals $\phi_r(\bm{r})$, $\phi_s(\bm{r})$:
\eqn{
	\hat V_j=\int_{Vol} d^3\bm{r} \frac{\hat \rho(\bm{r})}{\abs{\bm{s_j}-\bm{r}}}=\sum_{rs}\int_{Vol} d^3\bm{r} \frac{\rho_{rs}(\bm{r})\hat a^\dagger_r \hat a_s}{\abs{\bm{s_j}-\bm{r}}}=\sum_{rs}V_j^{rs}\hat a^\dagger_r \hat a_s.\label{eq:Vj}
}
Consequently, the molecule-nanoparticle coupling operator $\hat g_p$ defined in eq.(17 of the main text) can be obtained as:
\eqn{
	\hat g_{p}=\sum_{rs}q_{p,j}V_j^{rs}\hat a^\dagger_r \hat a_s=\sum_{rs}g_p^{rs}\hat a^\dagger_r \hat a_s.
}
with
\eqn{
	g_p^{rs}=\sum_{j}q_{p,j}V_j^{rs}
}
The coupling element between the $S_0\rightarrow S_1$ transition is then evaluated as:
\eqn{
	g^{bem,full}_p=\expect{S_0,1}{\sum_{rs}g^{rs}_p\hat a^\dagger_r \hat a_s \hat b^\dagger}{S_1,0}=\sum_{rs}\rho^{(S_0,S_1)}_{rs}g_p^{rs},\label{eq:gbemfull}
}
where we made use of the transition density matrix $\rho^{(S_0,S_1)}_{rs}$ to retrieve a similar formalism to the one proposed in refs.\cite{neuman:dipole,aizpurua:tautom}. The expression of $g^{bem,full}_p$ can be simplified by considering the  molecule as a point-dipole and retrieve therefore a Jaynes-Cummings-like\cite{jaynes:cav} 
description of the interaction between the nanoparticle and the molecule\cite{galego:cavity,mukamel:nadiab,herrera:spectra}. 
For a point-dipole molecule:
\eqn{
	g_p^{bem,dip}=\expect{S_0,1}{\sum_j q_{p,j}\frac{\left(\bm{s_j}-\bm{r_d}\right)\cdot\bm{\mu}^{S_0,S_1}}{\abs{\bm{s_j}-\bm{r_d}}^3}\hat b_p^\dagger}{S_1,0}=E_{q_{p}}\bm{\lambda_{q_p}}\cdot\bm{\mu}^{(S_0,S_1)}
\label{eq:gbemdip}
}
Here, $\bm{\mu}^{(S_0,S_1)}$ is the molecular transition dipole and  $E_{q_p}\bm{\lambda_{q_p}}$ (where $\bm{\lambda_{q_p}}$ is the polarisation unit vector), is the electric field associated to the apparent surface charges of the plasmonic mode under investigation $\bm{q_p}$:
\eqn{
	E_{q_p}\bm{\lambda_{q_p}}=\sum_j q_{p,j}\frac{\left(\bm{s_j}-\bm{r_d}\right)}{\abs{\bm{s_j}-\bm{r_d}}^3}
}
where $\bm{s_j}-\bm{r_d}$ is the distance between the tessera representative point $\bm{s_j}$ and the position of the point dipole $\bm{r_d}$. This expression directly connects to the typical $\hat H_{int}$ form used in approximated QED treatments where only the field-dipole term is retained, i.e.:

\eqn{
	\hat H^{dip}_{int}=\sum_p E_{1ph,p}\bm{\lambda_p} \cdot \bm{\hat \mu} \left(\hat b_p^\dagger+\hat b_p\right).\label{eq:hamdip}
}

For simple geometries, such as a nanosphere, the $E_{1ph,p}$ coefficient ($E_{q_p}$ when calculated from BEM) can be obtained analytically,\cite{delga:dipole} leading to another estimate of the coupling strength:

\eqn{
	g^{an,dip}_p=E_{1ph,p}\expect{S_0,1}{\hat g^{dip}_p \hat b_p^\dagger}{S_1,0}=E_{1ph,p}\bm{\lambda_p}\cdot \bm{\mu_{S_0,S_1}}
\label{eq:gandip}
}

The three simplified expressions introduced here, $g_p^{an,dip}$, $g_p^{bem,dip}$ and $g_p^{bem,full}$ are represented and compared in Figures \ref{fig:point_dip} and \ref{fig:comparison}. In the Results section of the main text we instead focus on applying this formalism to the case of realistic molecules, where we also provide polaritonic absorption spectra calculated with the full QED-CCSD-1 model and by considering polaritons obtained by a simple diagonalization of the system Hamiltonian in the $\ket{S_0,1}$ and $\ket{S_1,0}$ subspace, with $g_p^{bem,full}$ as the non-diagonal matrix element.

\section*{QED coupled cluster algorithm}

The results of this paper are calculated by coupling multiple advanced algorithms. Here we outline how the calculations have been performed and self-consistency reached.
From the geometry of the nanoparticle, the plasmon modes and corresponding charges and excitation energies are determined from the Drude-Lorentz parameters using the method outlined in the main text. This corresponds to determining the plasmonic basis. 
We write the plasmonic wave function as a linear combination of the modes $p$:
\begin{equation}
    \lvert P \rangle =
    \sum_{\vec{n}}  
    \prod_{p} (b^{\dagger}_p)^{n_p}
    \lvert 0 \rangle c_{\vec{n}}
\end{equation}
where we recognize the meaning of $\ket{0}$ as the plasmon vacuum and $\vec{n} = (n_1,n_2,\dots)$ is a vector collecting the plasmon occupation numbers in each mode\\

As presented in the main text, the electron-plasmon Hamiltonian reads:
\begin{equation}
    \hat H=\hat H_e +\sum_p \omega_p\hat b^\dagger_p \hat b_p +\sum_p \sum_j q_{p,j}\hat V_j\left(\hat b^\dagger_p + \hat b_p\right).\label{eq:qedh}
\end{equation}
The electrostatic potential operator $\hat V_j$ is defined in eq. \ref{eq:Vj}.
We now have a form of the Hamiltonian that resembles the Pauli-Fierz Hamiltonian presented by Haugland et al. \cite{haugland2020qedcc}. As detailed in the same reference, the non-correlated wave function is written as:
\begin{equation}
    \ket{R}=\ket{HF}\otimes\ket{P}. \label{eq:pwf}
\end{equation}
The plasmon charges $q_{p,j}$ and energy $\omega_P$ are input to QED-HF and the energy to the combined electron-plasmon system is minimized. The problem is solved in a self-consistent field until the coefficients of the electronic orbitals and plasmon number states reach self-consistency.
The result of this initial calculation yields the QED-HF reference state, and we report the main steps presented in ref. \cite{haugland2020qedcc} for our current Hamiltonian. The energy minimization with respect to the plasmonic coefficients is achieved by diagonalizing the Hamiltonian in eq. \ref{eq:qedh}. By applying the unitary coherent-state transformation:
\begin{equation}
    U(\mb{z})=\prod_p \exp(z_p(\hat b^\dagger_p-\hat b_p)),
\end{equation}
we obtain
\begin{equation}
   \bra{HF}U^\dagger(\mb{z}) \hat H U(\mb{z}) \ket{HF}=E_{HF}+\sum_p\left(\omega_p(\hat b^\dagger_p + z_p)(\hat b_p + z_p)+\sum_j q_{pj}\avg{\hat V_j}(\hat b_p+\hat b^\dagger_p + 2z_p )\right).
\end{equation}
Choosing the coherent state parameters $z_p$ as
\begin{equation}
    z_p=-\sum_j \frac{q_{pj}\avg{\hat V_j}}{\omega_p},
\end{equation}
we obtain in fact a diagonal Hamiltonian in the plasmonic number states (of the coherent state-transformed plasmon basis):
\begin{equation}
   \bra{HF}U^\dagger(\mb{z}) \hat H U(\mb{z}) \ket{HF}=E_{HF}+\sum_p\left(\omega_p\hat b^\dagger_p \hat b_p-\frac{1}{\omega_P}\sum_j q_{pj}\avg{\hat V_j}\sum_k q_{pk}  \avg{\hat V_k} \right).
\end{equation}
The expression in this equation is then minimized with respect to the electronic degrees of freedom. 
In each iteration of the minimization procedure we update the $\ket{HF}$ orbitals and use them to compute $\avg{\hat V_j}$. The Fock matrix for the orbital optimization now includes the plasmon-molecule interaction:
\begin{equation}
    F_{rs}=F^e_{rs} - \sum_p\left( \frac{2}{\omega_p} \sum_{j} q_{pj} \avg{\hat V_{j}} \sum_k q_{pk}\hat V_{k}^{rs} \right),
    \label{eq:fock}
\end{equation}
where $r,s$ denote the orbitals and $\bm{F}^e$ is the standard Fock matrix for closed shell systems. The Fock matrix expression can be made physically more transparent by noticing that in view of eqs. 5 and 7 in the main text, it turns out that:
\begin{equation}
- \sum_p \frac{2}{\omega_p} q_{pj} q_{pk}=Q^{IEF}_{kj}(0)
\end{equation}
where $Q^{IEF}_{kj}(0)$ is the classical response matrix evaluated for  a static perturbation ($\omega = 0$). This leads to:
\begin{equation}
    F_{rs}=F^e_{rs} +\sum_{jk} Q^{IEF}_{kj}(0)\avg{\hat V_{j}} \hat V_{k}^{rs},
\end{equation}
Based on eq. 1 in the main text, $\sum_j Q^{IEF}_{kj}(0)\avg{\hat V_{j}}$ are the classical polarization charges $q_k^{HF}$ induced on the plasmonic nanostructure due to the dielectric response to the molecule-generated electrostatic potential at the HF level. The Fock operator reduces then to:
\begin{equation}
    F_{rs}=F^e_{rs} +\sum_{k} q_k^{HF} \hat V_{k}^{rs}.
\end{equation}
Like in classical implicit solvation theories\cite{mennucci:pcm}, the extra term in the Fock operator represents the interaction of the molecular electrons with the dielectric polarization induced in the metal nanostructure by the molecule itself. The iterative resolution of eq.\ref{eq:fock} provides self-consistency of the molecular wave function and nanostructure polarization, for the plasmonic modes included in the calculation. Seen in the perspective of quantized plasmon states, in the calculation of the QED-HF reference state, the plasmonic states act as a basis for the polaritonic calculation, hence we optimize the coefficients of the plasmonic states, each of which is associated with a fixed set of charges $q_{p,j}$. This is reminiscent of how atomic orbitals are not optimized in an electronic structure calculation, it is their coefficients that are optimized.\\

After the reference QED-HF state is found, we obtain the coupled plasmon-molecule Hamiltonian:
\begin{equation}
    \hat H = \hat H_e +\sum_p \left( \omega_p\hat b^\dagger_p \hat b_p
    + \sum_j q_{pj} \left(\hat V_j - \avg{\hat V_j} \right) (\hat b_p + \hat b^\dagger_p)
    + \frac{1}{\omega_p} \sum_j q_{pj} \avg{\hat V_j} \sum_k q_{pk}  \left(\avg{\hat V_k} - 2\hat V_k \right) \right)
\end{equation}
where the expectation value of $\hat V_j$ is calculated with the HF wave function. At this stage, the QED-CCSD-1 ansatz is applied and the coupled cluster projection equations are solved to find the ground state. The excitation operators defined in the next section introduce correlation between the molecular and the plasmonic excitations, as well as readjust the plasmonic polarization to the CCSD molecular wave function.\\

\subsubsection*{QED coupled cluster (QED-CCSD) operators}

Here, we provide explicit expressions for the coupled cluster operators. Starting from the molecule without plasmon, the wave function is chosen to be a single Slater determinant $\ket{\text{HF}}$ (Hartree-Fock). Electron correlation is introduced through the exponential of the cluster operator acting on the $\ket{\text{HF}}$ determinant,
\begin{equation}
    \ket{\text{CC}} = e^{\hat T} \ket{\text{HF}}.
\end{equation}
The cluster operator $\hat T$ is expressed as
\begin{equation} \label{eq:cluster_e}
    \hat T = \hat T_1 + \hat T_2 + \dots + \hat T_{N_e}
\end{equation}
where $T_1$ are linear combinations of single excitations, $T_2$ are linear combinations of double excitations and so on. More specifically, $T_1$ and $T_2$ is given by
\begin{align}
    &\hat T_1 = \sum_{ai} t_{ai} \sum_\sigma \hat a^\dagger_{a\sigma} \hat a_{i\sigma} 
    \\
    &\hat T_2 = \frac{1}{2} \sum_{aibj} t_{aibj} \sum_{\sigma\tau} \hat a^\dagger_{a\sigma} \hat a_{i\sigma} \hat a^\dagger_{b\tau} \hat a_{j\tau}
\end{align}
where $a,b$ and $i,j$ are indices refer to virtual and occupied orbitals respectively. The operators $\hat a^\dagger$, $a$ refer to the standard second quantization notation adopted in electronic structure theory.
The coupled cluster method is equivalent to the exact diagonalization in the limit where all excitations are included in $\hat T$. Instead of formulating exact diagonalization in a more complicated manner, one truncates the cluster operator to $\hat T_1+\hat T_2$ with the aim of reducing the computational complexity. This truncated approach is named CCSD, and for more details we refer to ref. \cite{Helgaker2000book}.

\noindent In QED-CCSD-1, the cluster operator is limited to one plasmon with one plasmon mode,
\begin{align} 
    \hat T &= \hat T_1 + \hat T_2 + \hat S^1_1 + \hat S^1_2 + \hat \Gamma^1.
\end{align}
Each term is a linear combination of excitation operators,
\begin{align} 
    &\hat S^1_1 = \sum_{ai} s_{ai} \sum_\sigma \hat a^\dagger_{a\sigma} \hat a_{i\sigma} \hat b_p^\dagger,
    \\
    &\hat S^1_2 = \frac{1}{2} \sum_{aibj} s_{aibj} \sum_{\sigma\tau} \hat a^\dagger_{a\sigma} \hat a_{i\sigma} \hat a^\dagger_{b\tau} \hat a_{j\tau} \hat b_p^\dagger,
    \\
    &\hat \Gamma^1 = \gamma_1 \hat b_p^\dagger.
\end{align}
Comparing CCSD and QED-CCSD-1 also includes the correlation contribution induced on the electronic states by the plasmonic mode without adding computational complexity: indeed, the QED-CCSD-1 has the same scaling with respect to the number of orbitals as in the CCSD case ($N_{\text{orb}}^6$), and is computationally feasible for small to medium-sized molecules. In the limit where the interaction $\hat g \rightarrow 0$, these two methods are equivalent.

As mentioned in the SI section "Simplified coupling models", in this work we also provide approximated results akin to the JC model, although with a Quantum Chemistry derived coupling.
The bilinear interaction is treated as a perturbation. Now one can use standard HF and CCSD instead of their QED equivalent and find the electronic structure independently from the plasmons, and then compute the bilinear interaction as a property, see Eq. \eqref{eq:gbemfull}.
In this case we form a basis from the ground state and a selected subset of excited electronic states together with the plasmon number states, and diagonalize the total Hamiltonian similarly to a Jaynes-Cummings model. The difference between the two methods being that in QED-CCSD, all singles and doubles states interact with the plasmons, but in Jaynes-Cummings CCSD, only a subset of states interact with the plasmon, usually through an approximate interaction.

\subsection*{Oscillator strengths}
In this study, we use EOM-QED-CCSD-1 to compute the energies and oscillator strengths.
The oscillator strengths $f$ are computed as
\begin{equation}
    f = \frac{2}{3}(E_e - E_g) |\langle e | \hat d_e + d_p (\hat b_p^\dagger + \hat b_p) | g \rangle |^2
\end{equation}
where $\hat d_e$ is the electronic dipole moment operator and $d_p$ is the plasmon dipole moment. Here $g$ and $e$ are the ground and excited states. The above expression is for symmetric theories, but since CCSD and QED-CCSD-1 are nonsymmetric, the expression is slightly modified,
\begin{equation}
    f = \frac{2}{3}(E_e - E_g) \langle \Lambda | \hat d_e + d_p (\hat b_p^\dagger + \hat b_p) |R\rangle\langle L| \hat d_e + d_p (\hat b_p^\dagger + \hat b_p) | CC \rangle.
\end{equation}
Here $|R\rangle$ and $\langle L|$ are the right and left excited states of coupled cluster, $\langle\Lambda|$ is the left ground state and $|CC\rangle$ is the right coupled cluster ground state.

\subsection*{Supplementary Numerical Results}
\subsection*{Numerical comparison of simplified plasmon-molecule coupling models}

\begin{figure*}[ht!] 
	\centerline{\makebox[1.0\paperwidth][c]{
		\includegraphics[width=1\linewidth]{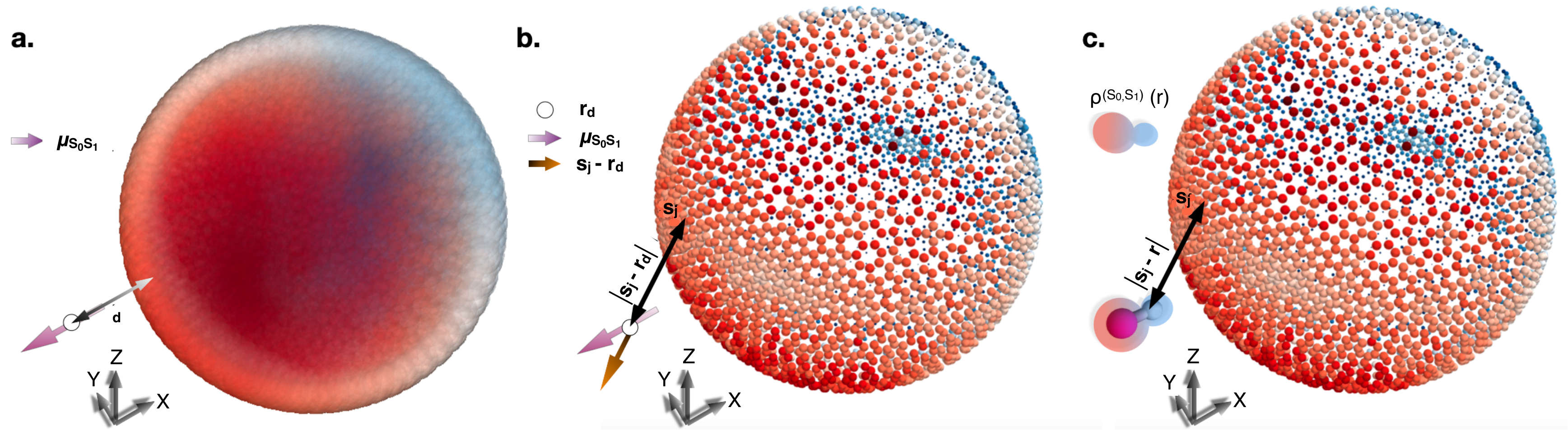}}}
	\caption{\textbf{Dipolar plasmonic mode of a spherical nanoparticle
	(diameter=10 nm) interacting with a molecule|} \textbf{a)} The molecule
	is represented as a point dipole distant $\bm d$ from the nanoparticle
	surface, while the nanoparticle is taken as a continuous Drude-Lorentz
	metal. The computed coupling is analytical, following ref.\cite{delga:dipole}.
	\textbf{b)-c)} The plasmonic modes are computed at Q-PCM-NP diagonal
	level. In Panel \textbf{b)}, the molecule is represented as a point
	dipole corresponding to the $S_0\rightarrow S_1$ transition dipole
	moment of the hydrogen fluoride molecule. The transition dipole moment $\mu_{S_0
	S_1}=1.695$ Debye is computed at EOM-CCSD level. 
	The EOM-CCSD calculations using the eT program\cite{tor} with a cc-pVTZ basis set.
	In Panel \textbf{c)} The molecule is represented with its
	$\rho^{(S_0,S_1)}(\bm{r})$ electronic transition density computed at
	CCSD level. The plasmon-molecule interactions are accounted 
	via eq. \ref{eq:g_p} and eq. \ref{eq:hamdip}.}\label{fig:point_dip}
\end{figure*}

\noindent Here, we will compare the coupling between a spherical nanoparticle and a molecule obtained in three different ways of increasing accuracy, as sketched in Figure \ref{fig:point_dip}:
\begin{itemize}
	\item{Coupling between analytical spherical plasmonic modes and a point-dipole molecule (Figure \ref{fig:point_dip}a), $g_p^{an,dip}$. This approach was proposed by Delga \textit{et al.}\cite{delga:dipole}.}
	\item{Coupling between the plasmonic modes computed at Q-PCM-NP level and a point-dipole molecule (Figure \ref{fig:point_dip}b), $g_p^{bem,dip}$. This is done to directly compare with the analytical results.}
	\item{Coupling between the plasmonic modes computed at Q-PCM-NP level and a realistic molecule, computed at EOM-CCSD level (Figure \ref{fig:point_dip}c), $g_p^{bem,full}$. This is a central feature of this work.}
\end{itemize}
As a test molecule, we take hydrogen fluoride due to its simple structure and longitudinal dipolar $S_0\rightarrow S_1$ transition. The dipolar behaviour of this transition allows us to compare the coupling computed with the realistic molecule (Figure \ref{fig:point_dip}c) to the other cases where the molecule is approximated as a point-dipole (Figures \ref{fig:point_dip}a and \ref{fig:point_dip}b).\\ 

We use a 5 nm radius spherical nanoparticle, the same one shown in Figure \ref{fig:point_dip}b. To directly compare the results of our method with the analytical model of A. Delga \textit{et al.}\cite{delga:dipole}, we define $l$ as the angular momentum of the quantized spherical nanoparticle, recalling that the plasmonic modes are $2l+1$ degenerate.
We compute the analytical coupling values $g^{an}_l$ by grouping and summing the degenerate plasmonic modes by angular momentum to remove potential artifacts due to the spatial orientation of the plasmonic modes. Following the derivation presented in ref.\cite{delga:dipole}, we recast the analytical coupling presented in eq. \ref{eq:gandip} into: 
\eqn{
	\left(g^{an,dip}_l\right)^2=\frac{\sqrt{l}(l+1)^2\mu^2_{S_0,S_1}}{2} \sqrt{\frac{\Omega_P^2}{2l+1}} \frac{r_{s}^{(2l+1)}}{d^{(2l+4)}},\label{eq:gandip2}
}
where $l$ is the angular momentum associated with the nanoparticle plasmonic modes manifold (three dipolar modes, five quadrupolar modes and so on) and $\Omega_P$ is the plasma frequency appearing in Drude-Lorentz, which determines $\omega_p$ (eq. 7 of the main text). Here $r_s$ and $d$ are respectively the radius of the sphere and the distance of the point dipole from the nanoparticle surface.\\

\begin{figure*}[ht!] 
	\centerline{\makebox[1.0\paperwidth][c]{
		\includegraphics[width=1\linewidth]{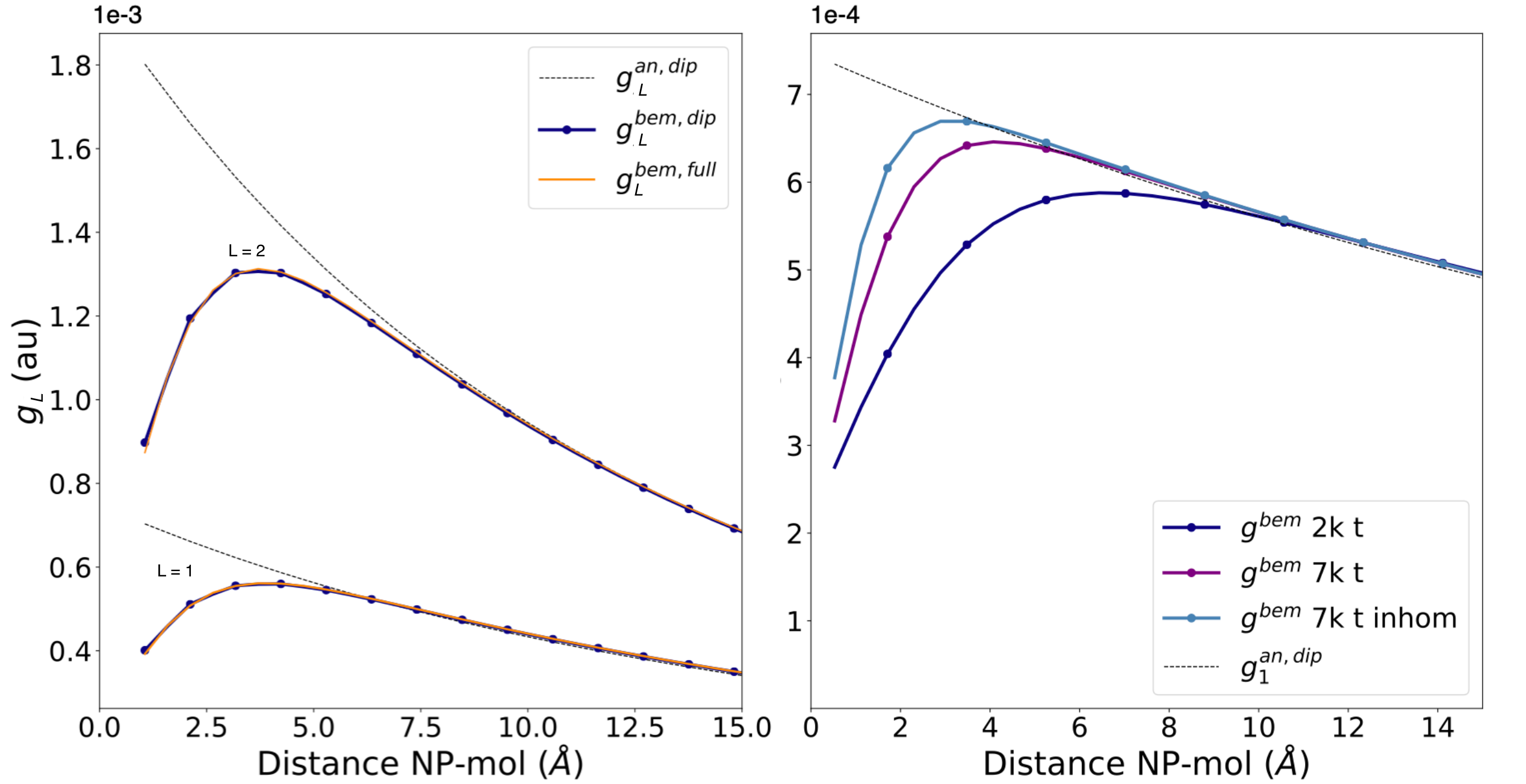}}}
	\caption{\textbf{Interaction between a molecule and the $\bm{\ell=1}$ (dipolar), $\bm{\ell=2}$ (quadrupolar) plasmonic modes of a nanosphere (diameter=10 nm)|} \textbf{a} Comparison 
	between the analytical value reported in (dashed black line)
	with the $g^{bem,dip}_\ell$ and $g^{full,dip}_\ell$. \textbf{b} Interaction element for the $\ell=1$ for the same case. The number of tesserae is increased from two thousands (labelled as 2k t, blue line) to seven thousands (labelled as 7k t, purple line). An even closer agreement between the analytical model and the BEM-based model is obtained by keeping seven thousands tesserae and making the mesh denser close to the molecule(labelled as 7k t inhom, light blue line).}\label{fig:comparison}
\end{figure*}

\noindent To directly compare to the analytical $(g^{an,dip}_l)$ value, we need to compute the interaction with the same modes with our method, hence taking into account the degeneracy of the sphere. To this aim, we sum the squared values of $g^{bem,dip}_p$ and $g^{bem,full}_p$ by angular momentum, respectively obtaining $(g^{bem,dip}_l)^2$ and $(g^{full,dip}_l)^2$.
In Figure \ref{fig:comparison}a,
we compare the values of $(g_l)^2$ for the three models. As plasmonic modes for the test, we take the $l=1$ (dipolar modes) and $l=2$ (quadrupolar modes) of the sphere. We use EOM-CCSD for an accurate description of the dipolar molecule (hydrogen fluoride).\\

\noindent A good agreement between the analytical model and the BEM variants is shown for
both the interaction elements with $\ell=1$ and $\ell=2$ up to $\sim$3 $\si{\angstrom}$ distance, comparable to the molecule-nanoparticle distance in plasmonic nanocavities and TERS setups\cite{baumberg:singlemol,vanhulst:nanocavity,aizpurua:tautom}. For real molecules, at shorter distances the analytical model would not be realistic anyway, as further effects like charge transfer would take place. The same behaviour is shown for both the variants of the coupling computed with the BEM plasmonic modes. The choice of a small dipole-like molecule such as hydrogen fluoride ensures that the realistic molecule is fully comparable to the point-dipole case. Indeed, no difference
is displayed when moving from point-dipole molecule ($g^{bem,dip}$) to the full
molecular description ($g^{bem,full}$). The discrepancy between the analytical model and our method is then due to the numerical accuracy of the BEM calculation, that we can relate to density of the tesserae on the nanoparticle surface. The effect becomes evident in Figure \ref{fig:comparison}b, where we show the
molecule-nanoparticle interaction for $\ell=1$
obtained by varying the number and density of tesserae (and hence of the ASC). An increase in the quality of the description is obtained by
increasing the number of tesserae from two thousands to seven thousands. A further
improvement is observed by inhomogeneously distributing the seven thousand tesserae
on the spherical surface, thickening the mesh close to the molecule. In the last case, the agreement between the analytical model and the BEM-based one holds down to $\sim 2.5$ $\si{\angstrom}$, which is even smaller than the typical non-bond distances
of 3.5 $\si{\angstrom}$. The excellent agreement between our method and the analytical results ensures that the plasmon-molecule interaction in the near-field is reliably described with our method.

\subsection*{Computational Details}

For the dimensions of the nanoparticles, please refer to Figure \ref{fig:bem}. The tassellation of the nanoparticles is performed with the GMSH\cite{gmsh} code by taking about 8k tesserae for the nanotip + support system, inhomogenoeusly distributed. We apply a linear-yet-steep gradient directed along the rotational axis of the nanotip and the support, thickening the tassellation towards the tip and on the top face of the support. By doing so, we obtain several hundreds of tesserae on the terminal part of the tip and ~2k on the top face of the support. For the sphere instead different tassellations are compared in the last section of the present Supporting Information.

The calculation on the classical nanoparticles and the quantised modes with the Q-PCM-NP is currently implemented in the software TDPlas\cite{pipolo:bem2}, available at: https://github.com/stefano-corni/WaveT\_TDPlas as a public repository. In our simulations for the nanoparticle + tip system we adopt the following Drude-Lorentz parameters:
\begin{itemize}
    \item $\Omega_p^2$=0.108 atomic units, corresponding to a bulk plasma frequency of 8.95 eV;
    \item $\gamma$=0.00077 atomic units (21 meV), corresponding to a decay rate of ;
    \item $\omega_0$=0, no natural frequency meaning we are operating in the Drude limit.
\end{itemize}
We chose the $\Omega_p$ such that the frequency of the tip mode represented in Figure 1c would be positioned between the two quasi-degenerate transitions of the free-base porphyrin, namely at 4.06 eV.
Such $\Omega_p$ is also very close to the Drude parameters for silver reported in ref.\cite{dlparams}. The $\gamma$ parameter is extracted from the same reference, always for the silver case.\\

 The electronic wave functions and molecular properties are calculated using EOM-CCSD (cc-pVDZ basis set).
 The calculations on porphyrin and PNA are performed using equation-of-motion QED-CCSD-1 with an aug-cc-pVDZ basis set, except for the Jaynes-Cumming calculations (JC), which are performed using transition densities from EOM-CCSD.
 All calculations using QED-CC in this paper are performed using a local branch of the eT program.\cite{tor} The branch is expected to be released in the near future.

\subsection*{Many-state Jaynes-Cummings}
\begin{figure}[!htpb]
    \centering
    \includegraphics[width=0.8\textwidth]{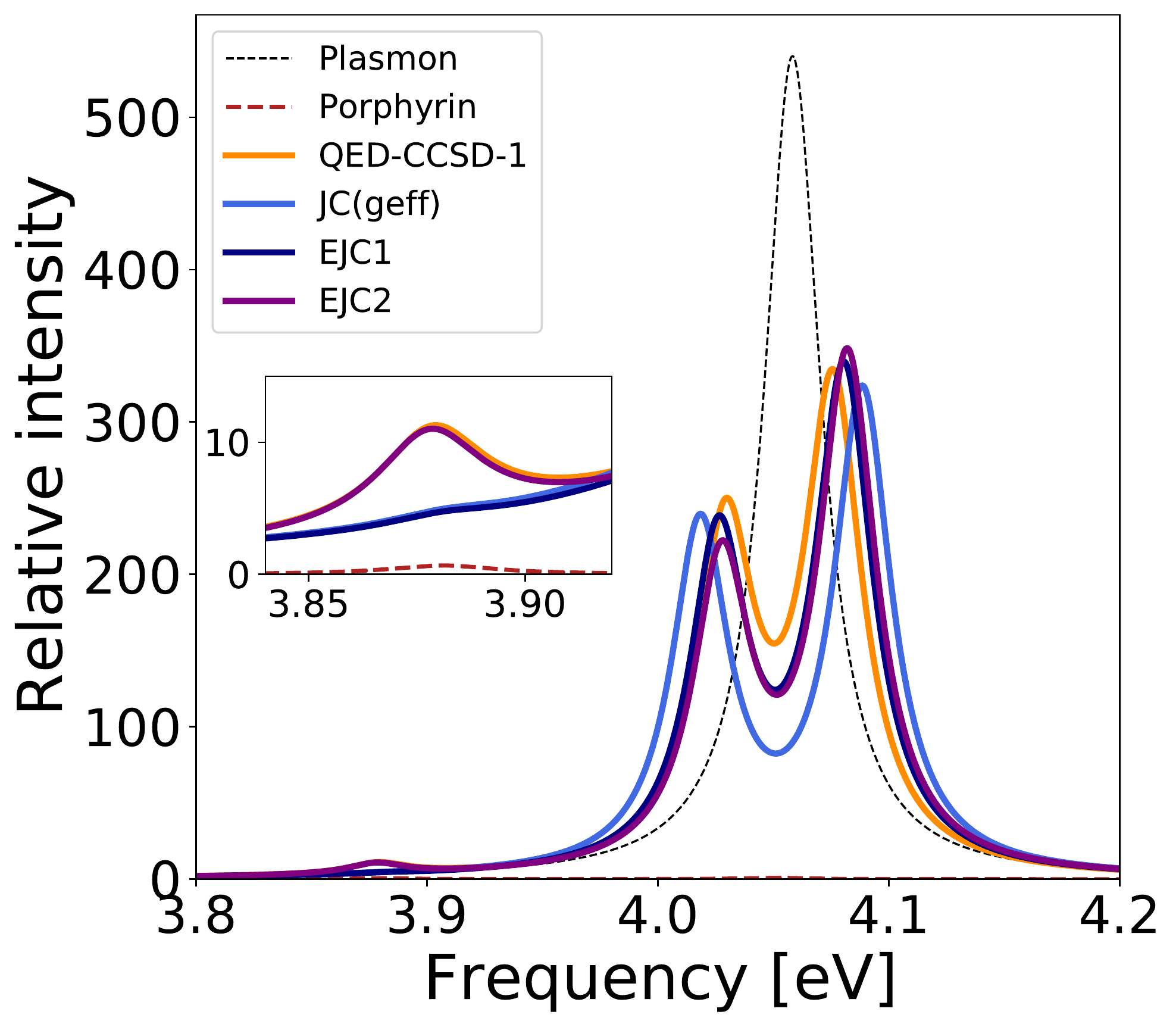}
    \caption{Porphyrin interacting with a plasmonic nanotip with transition energy $\omega=4.06$ eV (see Fig 2a). Oscillator strengths (intensity) are for porphyrin displaced 6 {\AA} along $x$ and 6 {\AA} along $y$ (see Fig 2b).
    Intensities are relative to porphyrin's strongest transition. The inset shows the same plot with only porphyrin.  The Rabi splittings denoting JC are calculated using the two-state JC with $g_{eff}$ and three-state (dark blue, EJC1 and purple, EJC2) Extended JC model with the excited state at $4.05$ eV. The effective coupling for JC is $g_{\text{eff}}=\sqrt{g_1^2+g_2^2}$. The EJC1 includes the ground state, the 4.05 eV and the 4.07 eV transition. EJC2 includes the ground state, the 3.88 eV and the 4.05 eV transitions. The comparison of the models in the region around 3.88 eV is displayed in the inset.}
    \label{fig:many_jc}
\end{figure}
The Jaynes-Cummings model used in the main text is defined by the Hamiltonian, $\hat H=\hat H_e + \hat H_{p} + \hat g_p$, in the basis of approximate eigenfunctions of $\hat H_e$ and eigenfunctions of $\hat H_{p}$ and applying the rotating-wave approximation. This leads to a 2-by-2 diagonalization problem to determine the dressed states and their energy,
\begin{equation}
\mathbf{H} = 
    \begin{bmatrix}
    E_1 & g_{eff} \\
    g_{eff} & E_2 \\
    \end{bmatrix},
\end{equation}
where $g_{\text{eff}}=\sqrt{g_1^2+g_2^2}$. When a three-state system is used for the molecule, then the dressed states are found by diagonalizing
\begin{equation}
\mathbf{H} = 
    \begin{bmatrix}
    E_1   & g_{1} & g_{2} \\
    g_{1} & E_2   & 0     \\
    g_{2} & 0     & E_3
    \end{bmatrix}.
\end{equation}
assuming that states 2 and 3 do not couple.
In Figure \ref{fig:many_jc} we show the different models applied to porphyrin interacting with a plasmonic nanotip. The effect of increasing the number of states has a small effect on the Rabi splitting and intensities. The three-state JC model, including the excited state with transition energy $3.88$ eV, also reveals a new peak in the spectrum, already shown by QED-CCSD.

\end{spacing}
\singlespacing
\clearpage
\providecommand{\latin}[1]{#1}
\makeatletter
\providecommand{\doi}
  {\begingroup\let\do\@makeother\dospecials
  \catcode`\{=1 \catcode`\}=2 \doi@aux}
\providecommand{\doi@aux}[1]{\endgroup\texttt{#1}}
\makeatother
\providecommand*\mcitethebibliography{\thebibliography}
\csname @ifundefined\endcsname{endmcitethebibliography}
  {\let\endmcitethebibliography\endthebibliography}{}